\documentclass[12pt]{article}
\usepackage{graphicx} 
\usepackage[margin=1in]{geometry}
\usepackage{multirow}%
\usepackage{subcaption}%
\usepackage[round,authoryear]{natbib}
\setcitestyle{authoryear,round}
\usepackage{amsmath}
\usepackage{xcolor}
\usepackage{booktabs,threeparttable}
\usepackage{comment}
\usepackage{adjustbox}
\usepackage{float}
\usepackage{amssymb}
\usepackage{url}

\title{Eigenvector Spatial Filters Nuclear Norm Matrix Completion with Application to Air Quality Data}
\author{Rodolfo Metulini}
\date{This version: August 24, 2026 \\ \hspace{2mm} \\ \small (Version submitted to "Spatial Statistics") \\}

\begin{document}

\maketitle

\abstract{Reliable reconstruction of missing observations in environmental panel datasets is essential for accurate exposure assessment and policy analysis. 
Traditional nuclear norm matrix completion methods effectively impute missing entries in low-rank matrices, yet often overlook the spatial dependence inherent to air quality processes. 
This paper introduces the Eigenvector Spatial Filters Nuclear Norm Matrix Completion (ESFNNMC) method, an extension of nuclear norm fixed-effects matrix completion that replaces unit-specific intercepts with a set of Moran-type eigenvectors capturing the dominant spatial dependence patterns implied by a spatial weights matrix. 
To estimate the model, we propose a block-coordinate descent algorithm, combined with soft-thresholded singular value decomposition and cross-validated regularization. 
Through comprehensive simulations varying missingness patterns, the level of spatial and temporal autocorrelation, and dimension, shape, and rank structure of the matrices, we demonstrate that ESFNNMC improves imputation accuracy when unit heterogeneity exhibits spatial structure, while remaining competitive under mild departures from this assumption.
Furthermore, it keeps the computational cost approximately unchanged.
The method is applied to impute missing entries in daily PM10 measurements in 64 monitoring stations in Lombardy, Italy, during the year 2021. 

\hspace{2mm}

}

\section{Introduction}

The reconstruction of missing values in panel datasets is a central problem in empirical research across the environmental sciences. 
Air quality monitoring, in particular, relies on high-frequency observations collected over spatially distributed monitoring networks, which are often incomplete due to sensor malfunction, calibration periods, or operational interruptions. 
Reliable imputation methods are therefore crucial for constructing accurate exposure measures and for supporting downstream analyses in environmental epidemiology, climate science, air quality management, and policy evaluation.

The problem of recovering missing entries in a matrix is commonly referred to as matrix completion \citep{laurent2009matrix}. 
The Netflix movie-rating challenge has become one of the canonical examples for matrix completion \citep{bennett2007netflix}.
Matrix completion via nuclear norm regularization has emerged as a powerful approach when the underlying data matrix can be well approximated by a low-rank structure \citep{fazel2002matrix, mazumder2010spectral,hastie2015statistical}. 
This framework is particularly suitable for environmental panel data, where temporal co-movements, seasonal dynamics, meteorological conditions, and shared latent factors often induce strong dependence across monitoring units.

Recent panel data matrix completion methods extend this idea by combining low-rank structures with fixed effects to recover missing outcomes and estimate counterfactuals \citep{athey2021matrix}. Recent applications have also demonstrated the potential of matrix completion for the counterfactual evaluation of environmental policies, including market-based tools and traffic restrictions \citep{biancalani2024impact, adam2026counterfactual}.
In this setting, the outcome matrix is decomposed into a latent low-rank component, additive unit and time effects, and an idiosyncratic error term. 
Such approaches provide a flexible alternative to standard two-way fixed effects and synthetic control methods, especially in the presence of time-varying unobserved heterogeneity \citep{bai2009panel}.
However, in environmental applications, unit effects are often not arbitrary: monitoring stations located in nearby or environmentally similar areas may exhibit systematically related behavior due to spatial dependence.

This observation motivates the methodological contribution of the present work. 
Building on the fixed-effects nuclear norm matrix completion estimator of \cite{athey2021matrix}, we propose a spatially structured extension in which unrestricted unit fixed effects are replaced by a parsimonious representation based on Eigenvector Spatial Filters (ESF). 
Rather than estimating a single effect for each spatial unit, we represent unit-specific heterogeneity as a linear combination of Moran eigenvectors extracted from a chosen spatial weights matrix. 
This allows the model to explicitly incorporate spatial autocorrelation while reducing the number of parameters to be estimated.

ESF techniques are grounded in Moran's $I$ and in the spectral properties of spatial weight matrices \citep{griffith2003spatial}. 
Following \cite{griffith2003spatial} and \cite{tiefelsdorf2007semiparametric}, Moran eigenvectors provide a multiscale decomposition of spatial dependence: leading eigenvectors describe broad geographical gradients, whereas subsequent eigenvectors capture progressively more localized regional and local spatial autocorrelation structures. 
ESF methods have been applied in several domains, including air pollution prediction \citep{Zhang2018Eigenvector}, species distribution analysis \citep{Diniz2005Modelling}, and international trade modeling \citep{patuelli2016space,metulini2018spatial}.

We refer to the proposed estimator as Eigenvector Spatial Filters Nuclear Norm Matrix Completion (ESFNNMC). 
The method combines three strands of literature: low-rank matrix completion via nuclear norm regularization, fixed-effects panel-data models, and eigenvector spatial filtering. 
Estimation is carried out through a block-coordinate descent algorithm that alternates between estimating the spatial coefficients, time effects, and the low-rank component, with the regularization parameter selected by cross-validation.

To evaluate the proposed method, we conduct an extensive simulation study varying matrix dimension and shape, missingness intensity, rank structure, temporal dependence, and the strength of spatial autocorrelation. 
Performance is assessed by comparing imputation accuracy through the Mean Absolute Percentage Error (MAPE) on validation entries. 
We then apply the method to daily PM$_{10}$ concentrations measured at 64 monitoring stations in Lombardy, Italy, during 2021. 
The empirical application illustrates how spatial filters provide an interpretable representation of the geographical structure of air pollution, while achieving predictive accuracy comparable to standard fixed-effects matrix completion, with improvements emerging particularly when spatial dependence is pronounced and temporal dependence is weak.

Overall, this paper contributes to the growing literature on spatial machine learning methods for environmental data reconstruction. 
It provides a computationally feasible and interpretable framework for recovering incomplete spatio-temporal matrices when the underlying data exhibit both low-rank latent structure and spatial dependence.

The remainder of the paper is organized as follows. 
Section~\ref{sec:methods} presents the methodological framework and the proposed ESFNNMC estimator. 
Section~\ref{sec:simulation} describes the simulation strategy, including robustness and sensitivity analyses. 
Section~\ref{sec:application} presents the application to PM$_{10}$ concentrations in Lombardy. 
Finally, Section~\ref{sec:conclusion} concludes and discusses limitations and directions for future research.
\section{Methods} \label{sec:methods}

\subsection{Nuclear Norm Matrix Completion via Soft-Impute}

We begin by considering a partially observed matrix $M \in \mathbb{R}^{n \times T}$ with associated observation mask $\Omega \subseteq \{1,\ldots,n\} \times \{1,\ldots,T\}$, where $\Omega_{it}=1$ if $M_{it}$ is observed and $0$ otherwise. 
Following \cite{mazumder2010spectral}, the nuclear norm matrix completion problem seeks a low-rank matrix $\hat{M}$ that best predicts the missing entries in $M$:
\begin{equation}\label{eq:mc}
\min_{\hat{M} \in \mathbb{R}^{n \times T}} \;
\frac{1}{2} 
\sum_{(i,t)\in \Omega} (M_{it} - \hat{M}_{it})^2 
\;+\; 
\lambda \, \|\hat{M}\|_*,
\end{equation}
where $\|\hat{M}\|_* = \sum_j \sigma_j(\hat{M})$ is the nuclear norm, i.e., the sum of the singular values of $\hat{M}$. 
The penalty, where $\lambda \geq 0$ is a regularization constant, controls the trade-off between fitting the known entries of the matrix and achieving a small nuclear norm, and so it encourages $\hat{M}$ to be low rank\footnote{The requirement of small nuclear norm is often related to getting a small rank of the obtained optimal solution of the optimization problem in Equation \ref{eq:mc}, which follows by geometric arguments similar to the ones typically adopted to justify how the classical LASSO (Least Absolute Shrinkage and Selection Operator) penalty term achieves effective feature selection in linear regression \citep{tibshirani1996regression}.}.

It was shown by \cite{mazumder2010spectral} that the optimization problem in Equation \ref{eq:mc} can be solved by the \emph{Soft-Impute} algorithm, which alternates between imputing missing entries and applying soft-thresholded singular value decomposition. 
Let $P_\Omega(\cdot)$ denote projection onto observed entries. 
First, ${\hat{M}}^{(0)}$ is initialized as ${ P}_{\Omega}({M}) \in \mathbb{R}^{n \times T}$ (i.e., missing entries are initially replaced by zero) and create a decreasing grid $\lambda_1 > \dots > \lambda_J$. 
For each value of $\lambda$, starting from the initial guess $\hat{M}^{(0)}$, Soft-Impute updates:
\begin{equation}
\hat{M}^{(k+1)} 
= 
\mathcal{S}_\kappa \!\left( 
P_\Omega(M) + P_{\Omega^c}\!\left(\hat{M}^{(k)}\right)
\right),
\end{equation}
where $\Omega^c$ represents the complement of $\Omega$, and $\mathcal{S}_\kappa(\cdot)$ denotes singular value soft-thresholding operator.
Given a matrix $H = U \Sigma V^\top$, with $\Sigma = $ diag$[\sigma_1, \cdots, \sigma_r]$ being the singular value decomposition of $H$, then:
\begin{equation}
\mathcal{S}_\kappa(H)
= 
U \, (\Sigma - \kappa I)_+ \, V^\top,
\qquad
(\cdot)_+ = \max\{0,\cdot\},
\qquad
\kappa=\frac{\lambda|\Omega|}{2}.
\end{equation}
By letting $\varepsilon > 0$ denote a selected tolerance, $k$ the step of the iteration, and $\|\cdot\|_F$ the Frobenius norm, this iterative procedure ends when $\frac{\|{\hat{M}}^{k + 1}-{\hat{M}}^{k} \|_F}{\|{\hat{M}}^{\rm k}\|_F} \leq \varepsilon$.
This approach shrinks singular values toward zero, removing weak latent factors and yielding a low-rank estimate.

\subsection{Fixed Effects Extension} \label{sec:FENNMC}

In many applications, especially related to environmental panel data and in economics, phenomena exhibit systematic differences between units (rows of the matrix) and between time periods (columns of the matrix) not captured solely by low-rank structure. 
\cite{athey2021matrix} therefore augment the nuclear norm matrix completion method defined above with unit and time fixed effects (hereinafter, FENNMC).
Estimation proceeds by solving:
\begin{eqnarray}\label{eq:mcfe}
\min_{{L} \in \mathbb{R}^{n \times T}, {u} \in \mathbb{R}^{n \times 1}, {v} \in \mathbb{R}^{T \times 1}} \;
\frac{1}{2}
\sum_{(i,t)\in \Omega}
\left( M_{it} - \hat{M}_{it} \right)^2
\;+\; 
\lambda \, \|L\|_*,
\\ 
\nonumber
\mbox{subject\,\,to}&&
M =
L + u {\bf 1}_{\emph{T}}^\top+{\bf 1}_{\emph{n}} v^\top\,,
\end{eqnarray}

where vectors $u$ and $v$ represent estimated unit and time effects, respectively, and ${\bf 1}_{\emph{n}}$ and ${\bf 1}_{\emph{T}}$ are column vectors consisting of $n$ entries and $T$ entries, respectively, all equal to $1$.
$\hat{L}$, $\hat{u}$ and $\hat{v}$ must be chosen to solve the above optimization problem.
It is worth noting that, in contrast to earlier formulations of the MC optimization problem \citep{mazumder2010spectral}, the nuclear norm $\|{L}\|_*$ is used in this optimization problem instead of the nuclear norm $\|{M}\|_*$. 
In other words, the estimated fixed effects $\hat{u} {\bf 1}_{\emph{T}}^\top+{\bf 1}_{\emph{n}} \hat{v}^\top$ are not regularized.
It is also worth noting that this specification requires the estimation of $n$ + $T$ fixed-effect parameters.

The fixed effects version is estimated through a Block Coordinate Descent (BCD) procedure that alternates between estimating the unit effects $u$, the time effects $v$, and the low-rank component $L$.
Block coordinate methods are widely used for optimization problems involving multiple parameter blocks, since they allow each
component to be updated conditionally on the others. 
General convergence results for block coordinate optimization are discussed, among others, by \cite{tseng2001convergence} and \cite{xu2013block}. 
In the present setting, each block update solves a convex subproblem and can be computed either in closed form or efficiently through
soft-thresholded singular value decomposition.

We initialize $L^{(0)}=0$, $u^{(0)}=0$, and $v^{(0)}=0$.
Conditional on the other parameter blocks, each update solves a convex problem and
admits a closed-form solution or can be efficiently computed.
\paragraph{1. Updating the unit effects.}
Given $L$ and $v$, the unit effects are estimated by least squares over the observed
entries:
\begin{equation}
u_i
=
\arg\min_{u_i}
\sum_{t:(i,t)\in\Omega}
\left(
M_{it}-L_{it}-v_t-u_i
\right)^2 . 
\end{equation}
The first-order optimality condition yields:
\begin{equation}
u_i
=
\frac{
\sum_{t:(i,t)\in\Omega}
\left(
M_{it}-L_{it}-v_t
\right)
}
{|\{t:(i,t)\in\Omega\}|}.
\end{equation}
Thus, each unit effect corresponds to the average row residual.
\paragraph{2. Updating the time effects.}
Given $L$ and $u$, the time effects are updated analogously:
\begin{equation}
v_t
=
\arg\min_{v_t}
\sum_{i:(i,t)\in\Omega}
\left(
M_{it}-L_{it}-u_i-v_t
\right)^2 ,
\end{equation}
which gives:
\begin{equation}
v_t
=
\frac{
\sum_{i:(i,t)\in\Omega}
\left(
M_{it}-L_{it}-u_i
\right)
}
{|\{i:(i,t)\in\Omega\}|}.
\end{equation}
\paragraph{3. Updating the low-rank component.}
Given $u$ and $v$, the low-rank matrix is updated through a Soft-Impute step. 
Define:
\begin{equation}
H=
P_{\Omega}
\left(
M-u\mathbf{1}_T^\top-\mathbf{1}_n v^\top
\right)
+
P_{\Omega^c}(L).
\end{equation}
Writing $H=U\Sigma V^\top$, with $\Sigma=\mathrm{diag}(\sigma_1,\ldots,\sigma_r)$,
the update is:
\begin{equation}
L \leftarrow S_\kappa(H).
\end{equation}
The algorithm iterates over these three steps until convergence.
This iterative procedure ends when $\frac{\|{{L}}^{k + 1}-{{L}}^{k} \|_F}{\|{{L}}^{\rm k}\|_F} \leq \varepsilon$, and the algorithm learns a latent low-rank structure $L$ while controlling for systematic unit and time differences.

\subsection{Eigenvector Spatial Filters Nuclear Norm Matrix Completion (ESFNNMC)}

Whereas fixed effects nuclear norm matrix completion (FENNMC) model defined in Equation \ref{eq:mcfe} decomposes the matrix M as:
\begin{equation}
M =
L +u {\bf 1}_{\emph{T}}^\top+{\bf 1}_{\emph{n}} v^\top\,,
\end{equation}
in our proposal, instead of estimating unit fixed effects vector $u \in \mathbb{R}^{n \times 1}$, we impose a spatial representation:
\begin{equation}
u = A\alpha,\qquad A\in\mathbb{R}^{n\times q},
\end{equation}
where $A$ is a matrix of exogenous spatial basis functions, where each column is one of the $q$ Moran eigenvectors of length $n$ to be externally selected, and $\alpha\in\mathbb{R}^q$ are unknown coefficients. 
Two are the main advantages with respect to FENNMC: 
\begin{itemize}
\item the number of parameters in vector $\alpha$ to be estimated is smaller than the number of parameters in vector $u$, since $q < n$;
\item the model provides an explicit and interpretable orthogonal representation of the spatial dependence structure among units.
\end{itemize}

We estimate $L$, $\alpha$, and $v$ by solving the following spatial filters fixed–effects matrix–completion problem:

\begin{eqnarray}\label{eq:esfmc}
\min_{{L} \in \mathbb{R}^{n \times T}, {\alpha} \in \mathbb{R}^{q \times 1}, {v} \in \mathbb{R}^{T \times 1}} \;
\frac{1}{2}
\sum_{(i,t)\in \Omega}
\left( M_{it} - \hat{M}_{it} \right)^2
\;+\; 
\lambda \, \|L\|_*,
\\ 
\nonumber
\mbox{subject\,\,to}&&
M =
\L +A\alpha\mathbf{1}_T^\top+{\bf 1}_{\emph{n}} v^\top\,.
\end{eqnarray}

The estimation strategy closely follows the FENNMC algorithm described above. 
The key difference is that the unrestricted unit effects $u$ are replaced by the spatial
representation $u=A\alpha$. 
We initialize $L^{(0)}=0$, $\alpha^{(0)}=0$, and $v^{(0)}=0$. 
Consequently, only the first block of the BCD procedure is modified, whereas the updates of $v$ and $L$ remain the same after replacing $u$ with $A\alpha$.
\paragraph{1. Updating the spatial coefficients.}
Given $L$ and $v$, the spatial coefficients are estimated by weighted least squares:
\begin{equation}
\alpha
=
\arg\min_{\alpha\in\mathbb{R}^q}
\sum_{i=1}^{n}
n_i
\left[
(A\alpha)_i
-
\frac{
\sum_{t:(i,t)\in\Omega}
\left(M_{it}-L_{it}-v_t\right)
}
{n_i}
\right]^2 ,
\end{equation}
where $n_i=|\{t:(i,t)\in\Omega\}|$. Equivalently,
\begin{equation}
(A^\top W A)\alpha=A^\top W y,
\end{equation}
with
\begin{equation}
W=\mathrm{diag}(n_1,\ldots,n_n),
\qquad
y_i=
\frac{
\sum_{t:(i,t)\in\Omega}
\left(M_{it}-L_{it}-v_t\right)
}
{n_i}.
\end{equation}
Unlike FENNMC, which estimates one unrestricted parameter for each unit, ESFNNMC
estimates a smaller set of coefficients associated with spatial eigenvectors.
\paragraph{2. Updating the time effects.}
Given $L$ and $\alpha$, the time effects are updated as in FENNMC, replacing $u_i$
with $(A\alpha)_i$:
\begin{equation}
v_t
=
\frac{
\sum_{i:(i,t)\in\Omega}
\left(
M_{it}-L_{it}-(A\alpha)_i
\right)
}
{|\{i:(i,t)\in\Omega\}|}.
\end{equation}
\paragraph{3. Updating the low-rank component.}
Given $\alpha$ and $v$, the low-rank matrix is updated through the same Soft-Impute
step:
\begin{equation}
H=
P_{\Omega}
\left(
M-A\alpha\mathbf{1}_T^\top-\mathbf{1}_n v^\top
\right)
+
P_{\Omega^c}(L),
\end{equation}
and, writing $H=U\Sigma V^\top$, $L\leftarrow S_\kappa(H)$.

The iterative procedure is repeated until:
$\frac{\|L^{k+1}-L^{k}\|_F}
{\|L^{k}\|_F} \leq \varepsilon$.

ESFNNMC can be interpreted as a spatially constrained version of FENNMC, where the unit-specific heterogeneity is restricted to lie in the subspace generated by the selected Moran eigenvectors.
When $A=I_n$, the two estimation approaches are equivalent; for general $A$, it constitutes a spatially structured FENNMC estimator that replaces idiosyncratic unit fixed effects with a spatial projection, improving recovery under spatially correlated heterogeneity.
The proposed specification assumes that a substantial part of the unit-specific heterogeneity is spatially structured and can therefore be represented within the subspace spanned by the selected Moran eigenvectors. 
Under this interpretation, the term $A\alpha$ captures the dominant geographical component of the unit effects, while residual non-spatial heterogeneity is absorbed by the low-rank component $L$ and the idiosyncratic error term. 
Consequently, ESFNNMC is expected to be most effective in settings where differences across units are primarily driven by spatial dependence rather than by purely unit-specific effects.

The resulting decomposition $M = L + A\alpha\mathbf{1}_T^\top + \mathbf{1}_n v^\top$ can be interpreted as the combination of three model components. 
The low-rank component $L$ captures latent spatio-temporal factors shared across multiple units and time periods, the term $A\alpha$ represents persistent spatial heterogeneity structured according to the spatial weights matrix, and the time effects $v$ account for temporal fluctuations common to all units. 
The relative importance of these components may provide useful insights into the mechanisms driving variability in the observed data and therefore constitutes an additional interpretative advantage of the proposed framework.

\subsubsection*{Convexity, convergence, and statistical considerations}

The proposed ESFNNMC estimator preserves the convex structure of the 
fixed-effects nuclear norm matrix completion problem. 
For a given spatial basis matrix $A$, the fitted values are affine functions of the unknown parameters $(L,\alpha,v)$. 
Since the squared-error loss and the nuclear-norm penalty are convex, the resulting objective function is jointly convex. 
Therefore, the restriction $u=A\alpha$ does not introduce non-convexity, but simply restricts the unit effects to the linear subspace spanned by the selected Moran eigenvectors.

The block-coordinate descent algorithm consequently operates on a jointly convex objective, with each block update solving a convex subproblem. 
Under standard regularity conditions, the objective decreases monotonically and limit points of the algorithm are global minimizers; see, for example, \citet{tseng2001convergence}. 
Thus, ESFNNMC preserves the computational properties of the fixed-effects formulation while replacing $n$ unrestricted unit effects 
with $q<n$ spatial coefficients.

Joint convexity does not, however, imply uniqueness of the decomposition into low-rank, spatial, and temporal components. 
In particular, spatial variation represented by $A\alpha$ may partly overlap with variation captured by the low-rank component $L$. 
The individual components should therefore be interpreted conditional on the selected spatial basis and the regularization imposed on $L$.

Existing results for nuclear-norm matrix completion provide a theoretical foundation for the proposed approach under suitable sampling and low-rank conditions \citep{candes2009exact,candes2010matrix,mazumder2010spectral,athey2021matrix}. 
Formal results on identification, consistency, finite-sample error, and recovery guarantees under the additional spatial restriction remain an important direction for future research.

\subsection{Cross-validation for $\lambda$}

A cross–validation is adopted to select the regularization parameter $\lambda$. 
Let $\Omega$ be randomly decomposed for each fold $k$ in training and validation sets in such a way $\Omega = \Omega^{\text{train}}_k \cup \Omega^{\text{val}}_k$. 
For each $\lambda$ on a decreasing logarithmic grid, the warm-started\footnote{According to the warm-started approach, instead of restarting from scratch for each $\lambda_j$, we use the solution at $\lambda_{j-1}$ as initialization.} block–coordinate algorithm is run on $P_{\Omega^{\text{train}}_k}(M)$, yielding estimates $(L^{(k)}_\lambda,\alpha^{(k)}_\lambda,v^{(k)}_\lambda)$. 
The validation error is:
\begin{equation}
\mathrm{RMSE}_k(\lambda)
=
\left(
\frac{1}{|\Omega^{\text{val}}_k|}
\sum_{(i,t)\in\Omega^{\text{val}}_k}
\big(
M_{it} - (L^{(k)}_{\lambda})_{it} - (A\alpha^{(k)}_{\lambda})_i - (v^{(k)}_{\lambda})_t
\big)^2
\right)^{1/2}
\end{equation}
We choose $\lambda^\ast = \arg\min_\lambda \frac1K\sum_{k=1}^K \mathrm{RMSE}_k(\lambda)$ and re–fit $(\ast)$ on all observed entries with warm–start initialization at $\lambda^\ast$, producing final estimates 
$\hat L, \hat\alpha, \hat v, \hat u = A\hat\alpha, \hat M = \hat L + \hat u\mathbf{1}_T^\top + \mathbf{1}_n\hat v^\top$.

\subsection{Weight matrix and the choice of the spatial eigenvectors} \label{sec:spatial}

Let $W \in \mathbb{R}^{n \times n}$ denote a spatial weights matrix of any type describing the neighborhood structure among the $n$ units. To extract spatial basis functions that capture structured dependence in $W$, we construct Moran–eigenvector spatial filters. First, $W$ is symmetrized as $\widetilde W = \frac{1}{2}(W+W^\top)$, and let $H = I_n - \frac{1}{n}\mathbf{1}_n\mathbf{1}_n^\top$ denote the centering matrix. 
Following a spectral approach to spatial filtering, we compute:
\begin{equation}
C \;=\; H\, \widetilde W\, H,
\end{equation}
and obtain its eigendecomposition:
\begin{equation}
C = V \Lambda V^\top,
\end{equation}
where $V=[v_1,\ldots,v_n]$ are orthonormal eigenvectors, and $\Lambda = \mathrm{diag}(\mu_1,\ldots,\mu_n)$ are the corresponding eigenvalues.
Eigenvectors associated with non-negligible positive eigenvalues are candidates for capturing positive spatial autocorrelation. 
In the first step, we retain indices:
\begin{equation}
\mathcal{I} = \{ j : \mu_j > \delta \},
\qquad \delta > 0,
\label{eq:retained}
\end{equation}
where $\delta$ must be a minimal positive value, and let $A_{\mathrm{full}} = [\,v_j : j \in \mathcal{I}\,]$.

Then, we assess the spatial autocorrelation strength encoded by each eigenvector in $A_{\mathrm{full}}$ by computing Moran's $I$ \citep{moran1950notes} for each column $a_j$ of $A_{\mathrm{full}}$, and we retain only eigenvectors with strictly positive spatial autocorrelation.
Eigenvectors associated with negative spatial autocorrelation may also contain meaningful spatial information. As emphasized by \cite{griffith2010detecting}, negative spatial autocorrelation represents a distinct form of spatial dependence that has received far less attention than its positive counterpart, despite its potential implications for spatial inference and model specification. 
Nevertheless, given the exploratory nature of this phenomenon and the predominance of positive spatial autocorrelation in the present context, we restrict our analysis to eigenvectors exhibiting positive spatial autocorrelation. 
A systematic investigation of the role and potential contribution of negatively autocorrelated eigenvectors is therefore left for future research.

Let $\widetilde W_{\mathrm{rs}}$ be the row-standardized version of $\widetilde W$, which is:
\begin{equation}
\widetilde W_{\mathrm{rs}} = \widetilde W D^{-1}, 
\qquad 
D = \mathrm{diag}(W\mathbf{1}_n),
\end{equation}
and $S_0 = \sum_{i}\sum_{t} \widetilde W_{\mathrm{rs},it}$. 
For each spatial eigenvector $a_j$, centered as $\tilde a_j = a_j - \bar a_j \mathbf{1}_n$, Moran's $I$ is computed as:
\begin{equation}
I_j 
= 
\frac{n}{S_0}
\frac{\tilde a_j^\top \widetilde W_{\mathrm{rs}} \tilde a_j}{\tilde a_j^\top \tilde a_j}.
\end{equation}
We retain only eigenvectors with spatial autocorrelation above zero:
\begin{equation}
\mathcal{I}_+ = \{ j \in \mathcal{I} : I_j > 0 \},
\end{equation}
we rank them by decreasing magnitude of $I_j$, and denote the ordered set by $\{I_{(1)}, I_{(2)}, \ldots\}$ with corresponding eigenvectors $\{a_{(1)}, a_{(2)}, \ldots\}$. Letting $I_{\mathrm{tot}} = \sum_{j \in \mathcal{I}_+} I_j$, we determine the number of spatial filters $q$ to be retained.
The threshold $\tau$ should be regarded as a hyperparameter controlling the complexity of the spatial representation. 
Larger values of $\tau$ retain a greater number of spatial filters, whereas smaller values lead to a more parsimonious specification\footnote{Generally, within a regression model framework, the choice of how many eigenvectors to retain is made through a stepwise procedure that includes only the significant eigenvectors \citep{chun2016eigenvector}. However, in this setting, we do not have a regression model with estimated coefficients associated with the explanatory variables, so such a procedure would not be a natural choice.}. 
We propose to determine the value of $q$ by an automatic selection rule ensuring that the chosen filters explain a specific percentage ($\tau$) of the total spatial autocorrelation signal, where $\tau$ can be determined using a sensitivity analysis or by cross-validation:
\begin{equation}
q
= 
\max\left\{
m : 
\frac{\sum_{j=1}^{m} I_{(j)}}{I_{\mathrm{tot}}}
\leq \tau
\right\}.
\end{equation}

The resulting spatial design matrix is:
\begin{equation}
A 
=
\big[\,a_{(1)}, \ldots, a_{(q)}\,\big] \in \mathbb{R}^{n \times q},
\end{equation}
whose columns serve as spatial basis functions capturing positive autocorrelation patterns among units. These eigenvectors are then used to model the eigenvector spatial filters component in the ESFNNMC model via the term $u = A\alpha$.

\section{Simulation Strategy} \label{sec:simulation}

In this section, we propose a simulation design strategy to evaluate the performance of the method in comparison with existing benchmarks, based on low-rank synthetic matrices with predefined levels of spatial and temporal autocorrelation, and with temporal regimes.
Details on the data generation process are in Section \ref{sec:design}.

\subsection{Low-Rank Spatio-Temporal Data Generating Process} \label{sec:design}

We generate a spatio-temporal matrix \(M \in \mathbb{R}^{n \times T}\) with a low-rank latent structure, spatial dependence across rows, temporal dependence across columns, and heterogeneous time-specific mean levels.
Let \(W \in \mathbb{R}^{n \times n}\) be a spatial weights matrix and \(\rho\) a spatial autoregressive parameter such that \((I - \rho W)\) is invertible. 
Define the spatial operator:
\begin{equation}
A = (I - \rho W)^{-1}. 
\end{equation}
The latent signal is constructed as:
\begin{equation}
M_0 = U S B^\top, 
\end{equation}
where \(U \in \mathbb{R}^{n \times r}\) contains unit-specific loadings, \(S \in \mathbb{R}^{r \times r}\) is diagonal with prescribed singular values, and \(B \in \mathbb{R}^{p \times r}\) contains latent temporal factors. 
Spatial dependence is introduced by filtering the latent signal:    
\begin{equation}
M_{\text{latent}} = A M_0.
\end{equation}
Temporal dependence is imposed by specifying each row of \(B\) as an AR(1) process:  
\begin{equation}
B_{t,\cdot} = \phi B_{t-1,\cdot} + \eta_t, \quad \eta_t \sim \mathcal{N}(0, I_r), \quad |\phi| < 1,
\end{equation}
with stationary initialization.
To allow for heterogeneous average levels across time, we introduce a vector of time effects \(\gamma \in \mathbb{R}^p\), possibly structured in regimes. 
The observed matrix is then defined as: 
\begin{equation}
M = M_{\text{latent}} + \mathbf{1}_n \gamma^\top + E,
\end{equation}
where \(\mathbf{1}_n\) is an \(n\)-dimensional vector of ones and \(E\) is a noise matrix with i.i.d.\ Gaussian entries. 
This specification induces column-specific mean shifts, so that different time periods may exhibit distinct average levels.
By construction, \(\operatorname{rank}(M_0) \le r\), and all rows of \(M_0\) lie in a common \(r\)-dimensional subspace of \(\mathbb{R}^p\), implying that a small number of latent temporal factors governs row trajectories.
We impose several feasibility and stability conditions. 
First, the spatial weight matrix must be conformable: $W \in \mathbb{R}^{n \times n}.$
Second, the target rank must satisfy $r \le \min(n, p).$
Third, the temporal parameter must satisfy the stationarity condition \(|\phi| < 1\). Finally, letting \(\lambda_{\max}(W)\) denote the spectral radius of \(W\), we require $|\rho| < \frac{1}{\lambda_{\max}(W)}$, which guarantees invertibility of \((I - \rho W)\).
To ensure numerical stability of the mean absolute percentage error (MAPE), we shift the simulated matrix by a positive constant. 
Specifically, we define:
\begin{equation}
M^{\star} = M + c,
\end{equation}
where \(c > -\min_{i,j} M_{ij} + \delta\) for some \(\delta > 0\), ensuring that $M^{\star}_{ij} \ge \delta > 0,  \forall i,j$.
This guarantees that the denominator in the MAPE remains bounded away from zero.

\hspace{2mm}

As an additional robustness check, we consider a data-generating process in which unit-specific heterogeneity is not spatially structured. 
The objective is to evaluate the behavior of ESFNNMC when its key assumption, namely that a substantial part of the differences across units is spatially structured, is not satisfied. 
In this setting, spatial dependence is removed and replaced by unrestricted unit-specific effects generated independently across units.
Specifically, the observed matrix is generated as:
\begin{equation}
M = U S B^\top + u\mathbf{1}_p^\top + \mathbf{1}_n\gamma^\top + E,   
\end{equation}
where $U S B^\top$ is the low-rank latent component, $u=(u_1,\ldots,u_n)^\top$ is a vector of unit-specific effects with $u_i \stackrel{i.i.d.}{\sim} N(0,\sigma_u^2),$ $\gamma$ denotes the vector of time effects, and $E$ is a Gaussian noise matrix. 
Temporal dependence is introduced through the same AR(1) specification adopted in the baseline simulation process. 
This alternative scenario preserves the low-rank and temporal structure of the data while removing spatial dependence, thereby providing a direct assessment of the robustness of ESFNNMC when the spatial-filter representation $u=A\alpha$ is not aligned with the true data-generating process.

\subsection{Simulation design}

We compute MAPE over B = 200 replications, for each combination of $\rho$ (spatial dependence parameter) and $\phi$ (temporal dependence parameter) in $(0,0.4,0.8)$, for missingness levels varying from $2\%$ to $25\%$, for small ($10 \times 10$) and large ($30 \times 30$) squared matrices, for long ($50 \times 10$) and wide ($10 \times 50$) rectangular matrices, with predefined rank $r$ ($r$ = 5 for small squared matrices and for wide rectangular matrices, $r=10$ for large squared matrices and long rectangular matrices) and $3$ time regimes.
We also record the optimal value of $\lambda$, the number of selected eigenvector spatial filters, and the total computational time.
The weight matrix $W$ is chosen to be a 0/1 binary matrix with $30\%$ of ones (i.e., connected pairs), while we choose $\tau = 0.90$ for the selection of the eigenvectors, after a sensitivity analysis we reported in Subsection \ref{sec:sens}. 

Both models are also tested without the inclusion of time fixed effects.

Computation times were measured on a system equipped with an 11th Gen Intel Core i5-1135G7 processor (2.40 GHz) and 16 GB of RAM. 
For a $10 \times 10$ matrix of rank 5 with $10\%$ missing entries, the average runtime per replication was 0.485 seconds for \text{ESFNNMC} and 0.432 seconds for \text{FENNMC}, indicating that the two methods exhibit comparable computational efficiency.

\begin{table}[!htbp]
\centering
\caption{ Median MAPE across methods for the $10\times 10$ matrix with rank $5$ and three regimes. For each missingness level, the best (lowest) median MAPE is in bold.}
\label{tab:mape_rank5_10x10}
\begin{adjustbox}{max totalheight=0.97\textheight,max width=\textwidth}
\begin{tabular}{lcccc}
\toprule
Missing (\%) & FENNMC (FE) & ESFNNMC (FE) & FENNMC (No FE) & ESFNNMC (No FE) \\
\midrule
\multicolumn{5}{c}{\textbf{Panel A: $\rho=0$, $\phi=0$}} \\
\midrule
2  & 4.936 & \textbf{3.875} & 7.494 & 4.549 \\
4  & 8.579 & \textbf{6.438} & 10.666 & 7.688 \\
6  & 9.373 & 8.143 & 12.710 & \textbf{7.837} \\
8  & 10.600 & \textbf{8.324} & 12.739 & 10.976 \\
10 & 12.514 & 11.328 & 15.085 & \textbf{10.495} \\
15 & 16.002 & 14.419 & 17.944 & \textbf{14.364} \\
20 & 19.051 & \textbf{15.735} & 21.100 & 17.272 \\
25 & 23.306 & \textbf{19.869} & 23.666 & 22.094 \\
\midrule
\multicolumn{5}{c}{\textbf{Panel B: $\rho=0$, $\phi=0.4$}} \\
\midrule
2  & 4.627 & \textbf{4.300} & 6.141 & 4.360 \\
4  & 6.574 & \textbf{6.368} & 9.504 & 7.181 \\
6  & 8.531 & 8.287 & 10.706 & \textbf{7.577} \\
8  & 8.651 & 8.648 & 11.193 & \textbf{9.706} \\
10 & 10.972 & \textbf{10.630} & 12.412 & 10.775 \\
15 & 15.373 & 14.244 & 16.340 & \textbf{13.555} \\
20 & 18.413 & 17.015 & 19.179 & \textbf{16.570} \\
25 & 25.743 & 23.467 & 22.805 & \textbf{21.303} \\
\midrule
\multicolumn{5}{c}{\textbf{Panel C: $\rho=0$, $\phi=0.8$}} \\
\midrule
2  & 3.664 & 3.544 & 3.851 & \textbf{3.292} \\
4  & 4.370 & 5.781 & 5.029 & \textbf{4.950} \\
6  & 7.055 & 7.583 & 6.511 & \textbf{6.189} \\
8  & \textbf{6.246} & 8.339 & 7.029 & 6.621 \\
10 & 8.998 & 9.423 & 8.363 & \textbf{7.792} \\
15 & 11.630 & 12.170 & 11.148 & \textbf{10.948} \\
20 & 15.082 & 16.491 & 14.437 & \textbf{13.299} \\
25 & 20.396 & 22.225 & \textbf{17.345} & 18.300 \\
\midrule
\multicolumn{5}{c}{\textbf{Panel D: $\rho=0.4$, $\phi=0$}} \\
\midrule
2  & 4.495 & \textbf{4.081} & 6.698 & 4.162 \\
4  & 7.922 & \textbf{6.438} & 9.507 & 8.156 \\
6  & 8.208 & \textbf{7.634} & 10.645 & 8.803 \\
8  & 9.501 & \textbf{8.106} & 12.106 & 9.688 \\
10 & 11.078 & 10.959 & 12.755 & \textbf{10.568} \\
15 & 14.868 & \textbf{13.125} & 16.238 & 13.821 \\
20 & 16.548 & \textbf{14.701} & 19.430 & 15.087 \\
25 & 22.159 & \textbf{18.925} & 22.502 & 21.005 \\
\midrule
\multicolumn{5}{c}{\textbf{Panel E: $\rho=0.4$, $\phi=0.4$}} \\
\midrule
2  & 3.891 & 3.851 & 5.270 & \textbf{3.850} \\
4  & \textbf{6.270} & 6.494 & 7.532 & 6.440 \\
6  & 7.693 & 8.212 & 9.087 & \textbf{7.245} \\
8  & \textbf{8.413} & 8.752 & 10.463 & 9.275 \\
10 & 9.840 & 10.513 & 11.782 & \textbf{10.026} \\
15 & 14.470 & 13.677 & 15.228 & \textbf{13.237} \\
20 & 16.156 & 15.593 & 17.656 & \textbf{15.143} \\
25 & 22.001 & 21.703 & 21.218 & \textbf{20.947} \\
\midrule
\multicolumn{5}{c}{\textbf{Panel F: $\rho=0.4$, $\phi=0.8$}} \\
\midrule
2  & \textbf{2.948} & 3.740 & 3.295 & 3.110 \\
4  & \textbf{4.258} & 5.998 & 4.428 & 4.460 \\
6  & 6.229 & 7.778 & \textbf{6.095} & 6.222 \\
8  & \textbf{6.219} & 8.875 & 6.607 & 6.278 \\
10 & 8.789 & 9.491 & 7.809 & \textbf{7.715} \\
15 & 10.931 & 12.137 & \textbf{10.421} & 10.647 \\
20 & 14.077 & 16.778 & 13.284 & \textbf{12.788} \\
25 & 19.720 & 22.642 & 17.847 & \textbf{18.757} \\
\midrule
\multicolumn{5}{c}{\textbf{Panel G: $\rho=0.8$, $\phi=0$}} \\
\midrule
2  & 3.298 & \textbf{3.013} & 3.779 & 3.628 \\
4  & 5.221 & \textbf{4.367} & 5.534 & 5.801 \\
6  & 6.197 & \textbf{5.313} & 5.660 & 5.972 \\
8  & 6.838 & \textbf{5.646} & 7.631 & 7.400 \\
10 & 7.827 & \textbf{6.596} & 7.996 & 7.950 \\
15 & 11.743 & \textbf{9.882} & 12.659 & 11.176 \\
20 & 13.513 & \textbf{12.741} & 15.466 & 12.252 \\
25 & 18.151 & \textbf{15.780} & 20.569 & 16.661 \\
\midrule
\multicolumn{5}{c}{\textbf{Panel H: $\rho=0.8$, $\phi=0.4$}} \\
\midrule
2  & \textbf{2.756} & 2.811 & 3.060 & 3.230 \\
4  & 4.508 & 4.418 & \textbf{4.323} & 5.018 \\
6  & 5.408 & 5.703 & \textbf{5.112} & 5.350 \\
8  & \textbf{6.570} & 6.758 & 7.425 & 6.910 \\
10 & \textbf{7.057} & 7.628 & 7.784 & 7.759 \\
15 & 11.514 & 10.900 & 10.636 & \textbf{11.099} \\
20 & 14.602 & 13.732 & 14.479 & \textbf{12.413} \\
25 & 17.958 & \textbf{16.703} & 19.054 & 17.362 \\
\midrule
\multicolumn{5}{c}{\textbf{Panel I: $\rho=0.8$, $\phi=0.8$}} \\
\midrule
2  & 1.816 & 3.364 & 2.076 & \textbf{2.514} \\
4  & \textbf{3.066} & 4.527 & 3.144 & 4.203 \\
6  & \textbf{4.028} & 5.927 & 4.661 & 4.464 \\
8  & \textbf{5.008} & 6.923 & 5.407 & 5.178 \\
10 & 6.301 & 8.045 & \textbf{5.882} & 6.331 \\
15 & 9.281 & 10.410 & \textbf{8.840} & 9.416 \\
20 & 12.615 & 13.737 & \textbf{11.198} & 11.434 \\
25 & 16.845 & 19.184 & \textbf{15.506} & 15.755 \\
\bottomrule
\end{tabular}
\end{adjustbox}
\end{table}

\begin{table}[!htbp]
\centering
\caption{Median number of selected spatial filters. The medians are identical across all $(\rho,\phi)$ combinations.}
\label{tab:filters_rank5_10x10}
\begin{tabular}{lc}
\toprule
Missing (\%) & Median number of filters \\
\midrule
2  & 2 \\
4  & 2 \\
6  & 2 \\
8  & 2 \\
10 & 2 \\
15 & 2 \\
20 & 2 \\
25 & 2 \\
\bottomrule
\end{tabular}
\end{table}

\begin{table}[!htbp]
\centering
\caption{Median MAPE across methods (10$\times$50 matrix, rank 5, 3 time regimes). Best performance in bold.}
\label{tab:mape_rank5_10x50}
\begin{adjustbox}{max totalheight=0.97\textheight,max width=\textwidth}
\begin{tabular}{lcccc}
\toprule
 & FENNMC (FE) & ESFNNMC (FE) & FENNMC (No FE) & ESFNNMC (No FE) \\
\midrule

\multicolumn{5}{c}{\textbf{Panel A: $\rho=0$, $\phi=0$}} \\
\midrule
2  & 0.2460 & \textbf{0.2397} & 0.3221 & 0.9641 \\
4  & \textbf{0.4175} & 0.4422 & 0.5933 & 1.3424 \\
6  & 0.6808 & \textbf{0.6530} & 0.8755 & 1.6832 \\
8  & \textbf{1.0179} & 1.0569 & 1.2908 & 2.0062 \\
10 & \textbf{1.6667} & 1.7327 & 1.9536 & 2.6597 \\
15 & 3.1805 & \textbf{3.1173} & 3.3594 & 4.1144 \\
20 & 4.8596 & \textbf{4.6974} & 4.8275 & 5.7868 \\
25 & \textbf{6.1438} & 6.2652 & 6.5654 & 7.2650 \\

\midrule
\multicolumn{5}{c}{\textbf{Panel B: $\rho=0$, $\phi=0.4$}} \\
\midrule
2  & \textbf{0.2667} & 0.2917 & 0.3133 & 1.013 \\
4  & \textbf{0.4329} & 0.5389 & 0.5854 & 1.317 \\
6  & \textbf{0.6576} & 0.6895 & 0.8072 & 1.572 \\
8  & \textbf{1.1456} & 1.2596 & 1.3178 & 2.057 \\
10 & \textbf{1.7155} & 1.9363 & 1.9563 & 2.667 \\
15 & \textbf{3.2887} & 3.3141 & 3.2888 & 4.251 \\
20 & \textbf{4.9892} & 5.0610 & 5.0242 & 5.828 \\
25 & \textbf{6.3175} & 6.4399 & 6.3222 & 7.128 \\

\midrule
\multicolumn{5}{c}{\textbf{Panel C: $\rho=0$, $\phi=0.8$}} \\
\midrule
2  & \textbf{0.2757} & 0.3909 & 0.3313 & 1.019 \\
4  & \textbf{0.3607} & 0.6003 & 0.4834 & 1.260 \\
6  & \textbf{0.6971} & 1.0126 & 0.8149 & 1.609 \\
8  & \textbf{1.0997} & 1.4841 & 1.1867 & 1.968 \\
10 & \textbf{1.5381} & 2.0612 & 1.7501 & 2.568 \\
15 & 3.0308 & 3.9690 & \textbf{2.9844} & 4.096 \\
20 & 4.8956 & 5.8748 & \textbf{4.5194} & 5.910 \\
25 & 6.6526 & 7.4280 & \textbf{5.9676} & 7.517 \\

\midrule
\multicolumn{5}{c}{\textbf{Panel D: $\rho=0.4$, $\phi=0$}} \\
\midrule
2  & \textbf{0.2262} & 0.2463 & 0.3144 & 0.9491 \\
4  & \textbf{0.4044} & 0.4294 & 0.5323 & 1.2876 \\
6  & \textbf{0.6010} & 0.6033 & 0.7891 & 1.5740 \\
8  & \textbf{0.9127} & 0.9148 & 1.1184 & 1.9747 \\
10 & \textbf{1.5459} & 1.6476 & 1.7126 & 2.5403 \\
15 & 2.8700 & \textbf{2.8464} & 3.1836 & 3.8139 \\
20 & 4.1763 & \textbf{4.1703} & 4.2737 & 5.1072 \\
25 & 6.0758 & \textbf{5.8318} & 5.8667 & 7.1302 \\

\midrule
\multicolumn{5}{c}{\textbf{Panel E: $\rho=0.4$, $\phi=0.4$}} \\
\midrule
2  & \textbf{0.2443} & 0.2874 & 0.3310 & 1.008 \\
4  & \textbf{0.3787} & 0.4499 & 0.5650 & 1.264 \\
6  & \textbf{0.5518} & 0.6858 & 0.7452 & 1.578 \\
8  & \textbf{0.9775} & 1.1525 & 1.0619 & 1.999 \\
10 & \textbf{1.6896} & 1.8732 & 1.7906 & 2.642 \\
15 & 3.0196 & 3.0588 & \textbf{2.8929} & 3.865 \\
20 & \textbf{4.5412} & 4.5653 & 4.5651 & 5.319 \\
25 & 5.8673 & 5.9050 & \textbf{5.6477} & 6.722 \\

\midrule
\multicolumn{5}{c}{\textbf{Panel F: $\rho=0.4$, $\phi=0.8$}} \\
\midrule
2  & \textbf{0.2564} & 0.429 & 0.3349 & 1.037 \\
4  & \textbf{0.3695} & 0.592 & 0.4645 & 1.206 \\
6  & \textbf{0.5942} & 1.017 & 0.7116 & 1.498 \\
8  & \textbf{0.9542} & 1.366 & 1.0546 & 1.789 \\
10 & \textbf{1.4451} & 2.148 & 1.6085 & 2.403 \\
15 & 2.8988 & 3.596 & \textbf{2.5372} & 3.772 \\
20 & 4.5984 & 5.635 & \textbf{4.2004} & 5.264 \\
25 & 6.1411 & 6.925 & \textbf{5.4281} & 6.728 \\

\midrule
\multicolumn{5}{c}{\textbf{Panel G: $\rho=0.8$, $\phi=0$}} \\
\midrule
2  & \textbf{0.1876} & 0.2110 & 0.3107 & 0.9229 \\
4  & \textbf{0.2813} & 0.3133 & 0.4409 & 1.1796 \\
6  & \textbf{0.3809} & 0.4521 & 0.5307 & 1.2597 \\
8  & \textbf{0.7418} & 0.7720 & 0.8130 & 1.7444 \\
10 & \textbf{1.1420} & 1.2469 & 1.1304 & 2.1585 \\
15 & 2.0192 & 2.0837 & \textbf{1.9016} & 2.8205 \\
20 & 3.1701 & 3.1494 & \textbf{2.7854} & 3.9328 \\
25 & 4.6652 & 4.5799 & \textbf{4.0739} & 5.4479 \\

\midrule
\multicolumn{5}{c}{\textbf{Panel H: $\rho=0.8$, $\phi=0.4$}} \\
\midrule
2  & \textbf{0.1969} & 0.2611 & 0.3320 & 0.9503 \\
4  & \textbf{0.2778} & 0.3736 & 0.4479 & 1.1615 \\
6  & \textbf{0.4514} & 0.5957 & 0.5460 & 1.2655 \\
8  & \textbf{0.7394} & 0.9874 & 0.8033 & 1.6000 \\
10 & \textbf{1.1968} & 1.3858 & 1.1130 & 2.0743 \\
15 & 2.1659 & 2.2521 & \textbf{1.8271} & 2.8432 \\
20 & 3.1713 & 3.3184 & \textbf{3.0106} & 4.0941 \\
25 & 4.2224 & 4.2189 & \textbf{3.7714} & 5.0671 \\

\midrule
\multicolumn{5}{c}{\textbf{Panel I: $\rho=0.8$, $\phi=0.8$}} \\
\midrule
2  & \textbf{0.2203} & 0.3724 & 0.3305 & 0.9568 \\
4  & \textbf{0.2781} & 0.5164 & 0.4019 & 1.1188 \\
6  & \textbf{0.4662} & 0.8270 & 0.5263 & 1.2608 \\
8  & 0.7124 & 1.1686 & \textbf{0.6715} & 1.4643 \\
10 & 1.1162 & 1.6942 & \textbf{1.0160} & 2.0210 \\
15 & 1.8657 & 2.5893 & \textbf{1.5376} & 2.6397 \\
20 & 3.2335 & 4.0394 & \textbf{2.7214} & 4.0858 \\
25 & 4.6447 & 5.4356 & \textbf{3.5892} & 5.3933 \\
\bottomrule
\end{tabular}
\end{adjustbox}
\end{table}

\begin{table}[!htbp]
\centering
\caption{Median MAPE (50$\times$10 matrix, rank 10, 3 time regimes). Best performance in bold.}
\label{tab:mape_rank10_50x10}
\begin{adjustbox}{max totalheight=0.97\textheight,max width=\textwidth}
\begin{tabular}{lcccc}
\toprule
 & FENNMC (FE) & ESFNNMC (FE) & FENNMC (No FE) & ESFNNMC (No FE) \\
\midrule

\multicolumn{5}{c}{\textbf{Panel A: $\rho=0$, $\phi=0$}} \\
\midrule
2  & 8.490 & \textbf{7.297} & 11.47 & 8.489 \\
4  & 9.552 & \textbf{8.099} & 11.71 & 9.896 \\
6  & 11.480 & \textbf{9.551} & 13.34 & 11.441 \\
8  & 11.567 & \textbf{9.636} & 13.40 & 11.300 \\
10 & 12.338 & \textbf{10.205} & 14.28 & 12.235 \\
15 & 15.207 & \textbf{12.985} & 17.22 & 15.043 \\
20 & 16.798 & \textbf{14.439} & 18.18 & 15.818 \\
25 & 19.331 & \textbf{16.747} & 20.86 & 17.673 \\

\midrule
\multicolumn{5}{c}{\textbf{Panel B: $\rho=0$, $\phi=0.4$}} \\
\midrule
2  & 7.348 & \textbf{6.559} & 9.317 & 7.609 \\
4  & 8.516 & \textbf{7.776} & 10.087 & 8.676 \\
6  & 10.074 & \textbf{8.691} & 11.219 & 10.458 \\
8  & 9.948 & \textbf{9.146} & 11.398 & 10.071 \\
10 & 10.074 & \textbf{9.579} & 11.905 & 10.645 \\
15 & 13.355 & \textbf{12.160} & 14.405 & 12.942 \\
20 & 15.075 & \textbf{14.004} & 16.480 & 14.473 \\
25 & 17.567 & \textbf{16.355} & 18.345 & 16.791 \\

\midrule
\multicolumn{5}{c}{\textbf{Panel C: $\rho=0$, $\phi=0.8$}} \\
\midrule
2  & 4.861 & \textbf{4.842} & 5.621 & 5.160 \\
4  & 5.527 & \textbf{5.140} & 6.066 & 5.828 \\
6  & 6.471 & 6.480 & 7.395 & \textbf{6.791} \\
8  & 7.319 & \textbf{7.158} & 7.899 & 7.586 \\
10 & 6.810 & \textbf{6.876} & 7.606 & 7.407 \\
15 & 8.965 & \textbf{8.474} & 9.365 & 8.949 \\
20 & 10.052 & \textbf{9.982} & 10.391 & 10.023 \\
25 & 12.064 & 12.247 & \textbf{12.472} & 12.416 \\

\midrule
\multicolumn{5}{c}{\textbf{Panel D: $\rho=0.4$, $\phi=0$}} \\
\midrule
2  & 8.602 & \textbf{7.227} & 11.20 & 8.488 \\
4  & 9.539 & \textbf{8.124} & 11.66 & 9.587 \\
6  & 11.275 & \textbf{9.461} & 13.16 & 11.518 \\
8  & 11.437 & \textbf{9.748} & 13.52 & 11.174 \\
10 & 12.103 & \textbf{10.379} & 13.99 & 12.223 \\
15 & 14.889 & \textbf{12.883} & 17.29 & 14.465 \\
20 & 16.279 & \textbf{13.690} & 17.88 & 15.872 \\
25 & 18.039 & \textbf{15.771} & 20.05 & 17.423 \\

\midrule
\multicolumn{5}{c}{\textbf{Panel E: $\rho=0.4$, $\phi=0.4$}} \\
\midrule
2  & 7.353 & \textbf{6.853} & 9.049 & 7.485 \\
4  & 8.482 & \textbf{7.692} & 9.819 & 8.407 \\
6  & 10.252 & \textbf{9.057} & 11.619 & 10.286 \\
8  & 10.142 & \textbf{9.433} & 11.450 & 10.051 \\
10 & 10.184 & \textbf{9.622} & 11.588 & 10.218 \\
15 & 12.943 & \textbf{12.117} & 14.246 & 12.909 \\
20 & 15.112 & \textbf{13.692} & 16.093 & 14.238 \\
25 & 18.076 & \textbf{16.880} & 18.876 & 16.854 \\

\midrule
\multicolumn{5}{c}{\textbf{Panel F: $\rho=0.4$, $\phi=0.8$}} \\
\midrule
2  & 4.842 & 4.881 & 5.494 & \textbf{5.275} \\
4  & 5.476 & \textbf{5.279} & 5.951 & 5.793 \\
6  & 6.306 & 6.480 & 7.309 & \textbf{6.831} \\
8  & 7.302 & 7.306 & 7.746 & \textbf{7.560} \\
10 & 6.753 & 6.956 & 7.500 & \textbf{7.241} \\
15 & 8.724 & 8.628 & 9.174 & \textbf{8.973} \\
20 & 9.890 & 10.299 & 10.793 & \textbf{10.079} \\
25 & 11.813 & 12.025 & 12.055 & \textbf{12.151} \\

\midrule
\multicolumn{5}{c}{\textbf{Panel G: $\rho=0.8$, $\phi=0$}} \\
\midrule
2  & 7.645 & \textbf{6.708} & 9.339 & 8.236 \\
4  & 8.860 & \textbf{7.453} & 10.246 & 8.791 \\
6  & 10.580 & \textbf{9.196} & 12.751 & 10.673 \\
8  & 10.835 & \textbf{8.837} & 12.261 & 10.492 \\
10 & 11.571 & \textbf{9.973} & 13.533 & 11.463 \\
15 & 13.928 & \textbf{11.650} & 15.843 & 13.669 \\
20 & 14.626 & \textbf{12.593} & 16.588 & 14.166 \\
25 & 16.991 & \textbf{15.225} & 19.627 & 16.647 \\

\midrule
\multicolumn{5}{c}{\textbf{Panel H: $\rho=0.8$, $\phi=0.4$}} \\
\midrule
2  & 6.490 & \textbf{6.287} & 7.977 & 7.085 \\
4  & 7.597 & \textbf{7.254} & 8.674 & 7.689 \\
6  & 9.456 & \textbf{8.473} & 10.990 & 9.788 \\
8  & 9.917 & \textbf{8.971} & 11.282 & 9.599 \\
10 & 9.810 & \textbf{9.497} & 12.042 & 10.158 \\
15 & 12.700 & \textbf{11.589} & 14.097 & 12.543 \\
20 & 13.878 & \textbf{12.776} & 15.210 & 13.420 \\
25 & 15.537 & \textbf{14.585} & 17.674 & 15.304 \\

\midrule
\multicolumn{5}{c}{\textbf{Panel I: $\rho=0.8$, $\phi=0.8$}} \\
\midrule
2  & \textbf{4.687} & 4.844 & 5.421 & 5.104 \\
4  & \textbf{5.117} & 5.190 & 5.775 & 5.475 \\
6  & \textbf{6.293} & 6.420 & 7.025 & 6.510 \\
8  & \textbf{6.761} & 6.941 & 7.213 & 6.921 \\
10 & \textbf{6.806} & 6.895 & 7.496 & 7.148 \\
15 & \textbf{8.281} & 8.422 & 9.175 & 8.753 \\
20 & \textbf{9.442} & 9.759 & 10.842 & 9.546 \\
25 & \textbf{10.810} & 10.796 & 11.987 & 10.984 \\

\bottomrule
\end{tabular}
\end{adjustbox}
\end{table}

\begin{table}[!htbp]
\centering
\caption{Median MAPE across methods (30$\times$30 matrix, rank 10, 3 time regimes). Best performance in bold.}
\label{tab:mape_rank10_30x30}
\begin{adjustbox}{max totalheight=0.97\textheight,max width=\textwidth}
\begin{tabular}{lcccc}
\toprule
 & FENNMC (FE) & ESFNNMC (FE) & FENNMC (No FE) & ESFNNMC (No FE) \\
\midrule

\multicolumn{5}{c}{\textbf{Panel A: $\rho=0$, $\phi=0$}} \\
\midrule
2  & 0.1760 & \textbf{0.1745} & 0.2191 & 0.8988 \\
4  & \textbf{0.2075} & 0.2141 & 0.2571 & 0.9503 \\
6  & 0.2654 & \textbf{0.2617} & 0.3722 & 1.0433 \\
8  & 0.3352 & \textbf{0.3237} & 0.5134 & 1.1289 \\
10 & 0.4914 & \textbf{0.4804} & 0.8933 & 1.2765 \\
15 & 1.3720 & \textbf{1.2242} & 2.1123 & 2.0160 \\
20 & 3.1545 & \textbf{2.7805} & 4.1327 & 3.3132 \\
25 & 5.1721 & \textbf{4.7642} & 6.1412 & 5.3698 \\

\midrule
\multicolumn{5}{c}{\textbf{Panel B: $\rho=0$, $\phi=0.4$}} \\
\midrule
2  & \textbf{0.1740} & 0.1896 & 0.2157 & 0.8725 \\
4  & \textbf{0.2218} & 0.2372 & 0.2862 & 0.9746 \\
6  & \textbf{0.2764} & 0.3135 & 0.3636 & 1.0590 \\
8  & \textbf{0.3148} & 0.3580 & 0.4884 & 1.1398 \\
10 & \textbf{0.5094} & 0.5317 & 0.7640 & 1.2878 \\
15 & 1.3042 & \textbf{1.2554} & 1.8688 & 1.9217 \\
20 & 2.9590 & \textbf{2.8767} & 3.7775 & 3.2809 \\
25 & 4.8172 & \textbf{4.7390} & 5.5919 & 5.1703 \\

\midrule
\multicolumn{5}{c}{\textbf{Panel C: $\rho=0$, $\phi=0.8$}} \\
\midrule
2  & \textbf{0.1809} & 0.2271 & 0.2114 & 0.8536 \\
4  & \textbf{0.2242} & 0.2638 & 0.2596 & 0.9303 \\
6  & \textbf{0.2362} & 0.3176 & 0.2834 & 0.9694 \\
8  & \textbf{0.2990} & 0.4123 & 0.3768 & 1.0845 \\
10 & \textbf{0.4327} & 0.5009 & 0.5333 & 1.1790 \\
15 & \textbf{0.8895} & 1.2101 & 1.2311 & 1.6473 \\
20 & \textbf{2.2322} & 2.6032 & 2.7219 & 2.6392 \\
25 & \textbf{3.8552} & 4.2827 & 4.0995 & 4.1272 \\

\midrule
\multicolumn{5}{c}{\textbf{Panel D: $\rho=0.4$, $\phi=0$}} \\
\midrule
2  & 0.1739 & \textbf{0.1737} & 0.2134 & 0.9101 \\
4  & \textbf{0.2067} & 0.2223 & 0.2537 & 0.9696 \\
6  & \textbf{0.2613} & 0.2738 & 0.3766 & 1.0562 \\
8  & 0.3305 & \textbf{0.3214} & 0.5037 & 1.1305 \\
10 & \textbf{0.4846} & 0.4866 & 0.8013 & 1.3148 \\
15 & 1.3486 & \textbf{1.2202} & 1.9568 & 2.0028 \\
20 & 3.0595 & \textbf{2.7246} & 3.7916 & 3.2200 \\
25 & 5.2433 & \textbf{4.5845} & 5.8501 & 5.0347 \\

\midrule
\multicolumn{5}{c}{\textbf{Panel E: $\rho=0.4$, $\phi=0.4$}} \\
\midrule
2  & \textbf{0.1668} & 0.1894 & 0.2155 & 0.8864 \\
4  & \textbf{0.2108} & 0.2402 & 0.2676 & 0.9767 \\
6  & \textbf{0.2697} & 0.3187 & 0.3617 & 1.0348 \\
8  & \textbf{0.3251} & 0.3815 & 0.4437 & 1.1247 \\
10 & \textbf{0.5177} & 0.5833 & 0.7321 & 1.3192 \\
15 & \textbf{1.2403} & 1.3066 & 1.7736 & 1.9069 \\
20 & 2.8899 & \textbf{2.8288} & 3.7214 & 3.3343 \\
25 & 4.7743 & \textbf{4.7718} & 5.4394 & 5.0397 \\

\midrule
\multicolumn{5}{c}{\textbf{Panel F: $\rho=0.4$, $\phi=0.8$}} \\
\midrule
2  & \textbf{0.1745} & 0.2167 & 0.2166 & 0.8458 \\
4  & \textbf{0.2112} & 0.2527 & 0.2510 & 0.9250 \\
6  & \textbf{0.2311} & 0.3016 & 0.2851 & 0.9771 \\
8  & \textbf{0.3068} & 0.4029 & 0.3711 & 1.0695 \\
10 & \textbf{0.4277} & 0.5417 & 0.5488 & 1.1968 \\
15 & \textbf{0.8751} & 1.2107 & 1.1162 & 1.6482 \\
20 & \textbf{2.1977} & 2.5718 & 2.5261 & 2.6275 \\
25 & \textbf{3.8589} & 4.5700 & 4.0198 & 4.0927 \\

\midrule
\multicolumn{5}{c}{\textbf{Panel G: $\rho=0.8$, $\phi=0$}} \\
\midrule
2  & \textbf{0.1527} & 0.1593 & 0.2550 & 0.9427 \\
4  & \textbf{0.1836} & 0.1973 & 0.2807 & 1.0031 \\
6  & \textbf{0.2318} & 0.2510 & 0.3800 & 1.0774 \\
8  & \textbf{0.2721} & 0.2944 & 0.4223 & 1.1549 \\
10 & \textbf{0.4422} & 0.4523 & 0.6717 & 1.3273 \\
15 & \textbf{1.1392} & 1.1767 & 1.5451 & 1.9031 \\
20 & 2.7341 & \textbf{2.4481} & 3.0945 & 3.0379 \\
25 & 4.0261 & \textbf{3.6729} & 4.1964 & 4.1591 \\

\midrule
\multicolumn{5}{c}{\textbf{Panel H: $\rho=0.8$, $\phi=0.4$}} \\
\midrule
2  & \textbf{0.1488} & 0.1740 & 0.2508 & 0.9132 \\
4  & \textbf{0.1840} & 0.2110 & 0.2747 & 0.9905 \\
6  & \textbf{0.2232} & 0.2723 & 0.3644 & 1.0492 \\
8  & \textbf{0.2857} & 0.3523 & 0.4460 & 1.1607 \\
10 & \textbf{0.4503} & 0.5435 & 0.6391 & 1.3616 \\
15 & \textbf{1.0380} & 1.1636 & 1.3372 & 1.7645 \\
20 & \textbf{2.6410} & 2.6713 & 2.9295 & 3.0194 \\
25 & \textbf{3.9392} & 4.0078 & 4.3143 & 4.3347 \\

\midrule
\multicolumn{5}{c}{\textbf{Panel I: $\rho=0.8$, $\phi=0.8$}} \\
\midrule
2  & \textbf{0.1523} & 0.1918 & 0.2494 & 0.8648 \\
4  & \textbf{0.1897} & 0.2236 & 0.2731 & 0.9246 \\
6  & \textbf{0.2132} & 0.2811 & 0.3038 & 0.9771 \\
8  & \textbf{0.2601} & 0.3558 & 0.3817 & 1.0451 \\
10 & \textbf{0.4035} & 0.5104 & 0.5121 & 1.2166 \\
15 & \textbf{0.7815} & 1.0782 & 0.9249 & 1.5633 \\
20 & \textbf{1.8382} & 2.3786 & 1.8947 & 2.3363 \\
25 & 3.0980 & 3.6691 & \textbf{3.0806} & 3.5000 \\

\bottomrule
\end{tabular}
\end{adjustbox}
\end{table}

\hspace{2mm}

\subsection{Simulation results}

The main simulation results reported in Tables~\ref{tab:mape_rank5_10x10}, \ref{tab:mape_rank5_10x50}, \ref{tab:mape_rank10_50x10}, and \ref{tab:mape_rank10_30x30} provide consistent evidence on the comparative performance of the proposed ESFNNMC estimator relative to standard matrix completion approaches. Across all configurations, imputation accuracy, measured by median MAPE, deteriorates monotonically with the proportion of missing data, as expected. 
However, ESFNNMC generally outperforms FENNMC when moderate or strong spatial dependence is present, where the gains are persistent across all missingness levels. 
Additional summaries of the selected regularization parameters and the number of retained spatial filters are reported in Appendix~\ref{app:simulation_tables}.

The advantage of incorporating spatial filters is especially evident in squared matrices (both 10$\times$10 and 30$\times$30), where ESFNNMC with fixed effects achieves the lowest MAPE in most scenarios, highlighting its ability to exploit spatial autocorrelation efficiently. 

For 10$\times$50 rectangular matrices (wide), performance differences across methods are reduced, yet FENNMC with fixed effects remains competitive and sometimes outperforms ESFNNMC, suggesting that the presence of long and auto-correlated time series mitigates the relative contribution of spatial structure.

For 50$\times$10 rectangular matrices (long), ESFNNMC generally matches or outperforms FENNMC whenever moderate or strong spatial dependence is present, suggesting that our proposed method is particularly suitable in the presence of relatively short time series and a relatively large number of units in space.

Both methods (FENNMC and ESFNNMC), when time fixed effects are excluded, exhibit substantially higher errors, confirming the importance of controlling for both spatial and temporal heterogeneity. 

The selected regularization parameters, reported in Appendix~\ref{app:simulation_tables}, show a clear increasing pattern with higher missingness, indicating that stronger shrinkage is required in sparser settings. 
ESF-based models tend to select slightly lower or comparable $\lambda$ values relative to FENNMC, reflecting a more efficient representation of heterogeneity that is filtered out before the regularization. 
The number of selected spatial filters remains remarkably stable across all designs; see Table~\ref{tab:filters_rank5_10x10} for the $10\times10$ case and Appendix~\ref{app:simulation_tables} for the remaining matrix configurations. 
Across all simulation settings, the median number of selected filters is approximately 20--30\% of the number of spatial units, highlighting a large reduction in the number of parameters to be estimated.

Finally, computational times are comparable across methods, confirming that the inclusion of spatial filtering does not impose a significant additional burden. 
Overall, the results demonstrate that ESFNNMC provides improvements in imputation accuracy - particularly when spatial autocorrelation is present - without compromising computational feasibility.

\hspace{2mm}

As an additional robustness check, we consider a stress-test scenario in which unit-specific heterogeneity is not spatially structured. 
In this design, the spatial autoregressive component is replaced by unrestricted unit effects generated independently across units (see Section \ref{sec:design}). 
The purpose is to assess the behavior of ESFNNMC when the spatial filter restriction $u=A\alpha$ is not aligned with the data-generating process.
The results in Table~\ref{tab:nonspatial_fe_stress} show that, when the strength of the non-spatial unit effects increases, FENNMC tends to outperform ESFNNMC. 
This can be expected, since FENNMC estimates unrestricted unit effects, whereas ESFNNMC restricts unit heterogeneity to the subspace spanned by the selected Moran eigenvectors. 
Nevertheless, the differences remain moderate when the non-spatial unit effects are weak, suggesting that ESFNNMC is not severely penalized under mild departures from spatially structured heterogeneity. 
Overall, this stress test confirms that the advantage of ESFNNMC is most relevant when unit-specific heterogeneity is spatially structured, while FENNMC is preferable when such heterogeneity is purely idiosyncratic.

\begin{table}[!htbp]
\centering
\caption{Stress-test simulation with non-spatial unit fixed effects. Median MAPE for FENNMC and ESFNNMC for different strengths of unit effects ($\sigma_u$) and temporal dependence ($\phi$). 10\% and 20\% of missing values. 
Lower values indicate better performance.}
\label{tab:nonspatial_fe_stress}
\small
\begin{tabular}{ccrrrr}
\toprule
$\sigma_u$ & $\phi$ & 
\multicolumn{2}{c}{10\% missing} & 
\multicolumn{2}{c}{20\% missing} \\
\cmidrule(lr){3-4}\cmidrule(lr){5-6}
 &  & FENNMC & ESFNNMC & FENNMC & ESFNNMC \\
\midrule
0  & 0.0 & 12.514 & \textbf{11.328} & 19.051 & \textbf{15.735} \\
0  & 0.4 & 10.972 & \textbf{10.630} & 18.413 & \textbf{17.015} \\
0  & 0.8 & \textbf{8.998} & 9.423 & \textbf{15.082} & 16.491 \\
\midrule
5  & 0.0 & 12.866 & \textbf{12.011} & \textbf{18.424} & 18.982 \\
5  & 0.4 & \textbf{12.521} & 13.128 & 20.772 & \textbf{20.142} \\
5  & 0.8 & \textbf{9.760} & 11.184 & \textbf{14.734} & 18.029 \\
\midrule
15 & 0.0 & \textbf{12.706} & 14.815 & \textbf{19.491} & 22.333 \\
15 & 0.4 & \textbf{12.504} & 15.871 & \textbf{20.003} & 23.028 \\
15 & 0.8 & \textbf{9.732} & 13.007 & \textbf{15.142} & 19.518 \\
\bottomrule
\end{tabular}
\end{table}

\subsection{Sensitivity analysis on the hyperparameter $\tau$} \label{sec:sens}

The number of spatial filters included in the ESFNNMC model is determined through the cumulative spatial autocorrelation threshold $\tau$, which controls the proportion of the total positive Moran's $I$ signal explained by the selected eigenvectors.
To assess the sensitivity of the results to this choice, we replicated the simulation design using alternative thresholds $\tau \in \{0.80, 0.90, 0.95\}$. 
For computational convenience, we report here a representative scenario corresponding to square $10 \times 10$ matrices with rank $r = 5$ and three temporal regimes. 
As in the main simulation study, results are based on $B = 200$ replications.
Table \ref{tab:sensitivity_tau} reports the median MAPE, the median number of selected spatial filters, and the median value of the optimal regularization parameter $\lambda$ for each threshold. 
Globally, the results show that the performance of ESFNNMC is only
marginally affected by the choice of $\tau$, suggesting that the proposed procedure is robust to moderate variations in the amount of spatial autocorrelation retained.
As expected, larger values of $\tau$ lead to the inclusion of a greater number of spatial filters, although the corresponding gains in prediction accuracy remain limited.
These findings support the use of $\tau = 0.90$ as a parsimonious compromise between model complexity and predictive performance.

\begin{table}[ht]
\centering
\caption{Sensitivity analysis with respect to the cumulative Moran's $I$ threshold for spatial filter selection. Results refer to the $10\times10$ matrix with rank $5$, three temporal regimes, $\rho=0.4$, and $\phi=0.4$. Reported values are medians
over $B=200$ replications.}
\small
\begin{tabular}{cccccccccc}
\toprule
&
\multicolumn{3}{c}{$\tau=0.80$}
&
\multicolumn{3}{c}{$\tau=0.90$}
&
\multicolumn{3}{c}{$\tau=0.95$} \\
\cmidrule(lr){2-4}
\cmidrule(lr){5-7}
\cmidrule(lr){8-10}
Missing (\%)
& MAPE & $\lambda$ & Filters
& MAPE & $\lambda$ & Filters
& MAPE & $\lambda$ & Filters \\
\midrule
2  & 3.723  & 0.2365 & 2 & 3.851 & 0.2751 & 2 & 4.461 & 0.2852 & 3 \\
4  & 5.834  & 0.3177 & 2 & 6.494 & 0.3486 & 2 & 6.763 & 0.3767 & 2 \\
6  & 8.111  & 0.3148 & 2 & 8.212 & 0.3473 & 2 & 8.130 & 0.3501 & 2 \\
8  & 8.306  & 0.3426 & 2 & 8.752 & 0.3066 & 2 & 9.275 & 0.3480 & 2 \\
10 & 10.065  & 0.3544 & 2 & 10.513 & 0.3990 & 2 & 10.542 & 0.3913 & 2 \\
15 & 13.206 & 0.5319 & 2 & 13.677 & 0.6082 & 2 & 13.782 & 0.6035 & 2 \\
20 & 15.665 & 0.7771 & 2 & 15.593 & 0.7313 & 2 & 16.154 & 0.7313 & 2 \\
25 & 21.286 & 0.9730 & 2 & 21.703 & 1.0086 & 2 & 21.468 & 0.9833 & 2 \\
\bottomrule
\end{tabular}
\label{tab:sensitivity_tau}
\end{table}

Across all missingness levels, median MAPE values remain remarkably stable when $\tau$ varies from 0.80 to 0.95. 
Similarly, the optimal regularization parameter $\lambda$ exhibits only minor variations, while the number of selected filters remains small and relatively constant. 
These results suggest that the proposed estimator is robust to moderate changes in the amount of spatial autocorrelation retained.

\section{Application to Air Quality} \label{sec:application}

\subsection{Data}
The analysis relies on the \textit{Agrimonia} dataset, a high–resolution environmental monitoring resource publicly released in the journal \textit{Scientific Data}. 
The \textit{Agrimonia} dataset \citep{fasso2023agrimonia} provides a harmonized collection of environmental, meteorological, and socio-economic variables for Northern Italy. 
It integrates measurements from the \textit{Regional Environmental Protection Agencies} (ARPA), \textit{ERA5-Land} reanalysis, and \textit{ISTAT} census databases. 
The dataset covers the period 2013--2021 with daily temporal resolution and station-level spatial granularity. 
Core variables include air pollutants (PM$_{10}$, PM$_{2.5}$, NO$_2$, O$_3$), meteorological indicators (temperature, humidity, wind speed, precipitation), and ancillary socio-economic descriptors. 
Data are harmonized and gap-filled using spatial and temporal interpolation procedures ensuring consistency across stations and years. 
The resulting integrated framework supports reproducible analyses of air quality dynamics, environmental inequalities, and spatial-temporal dependence in pollution exposure across Northern Italy.

\begin{figure}[htbp!]
\begin{subfigure}{0.49\textwidth}
\centering
\includegraphics[width=\textwidth]{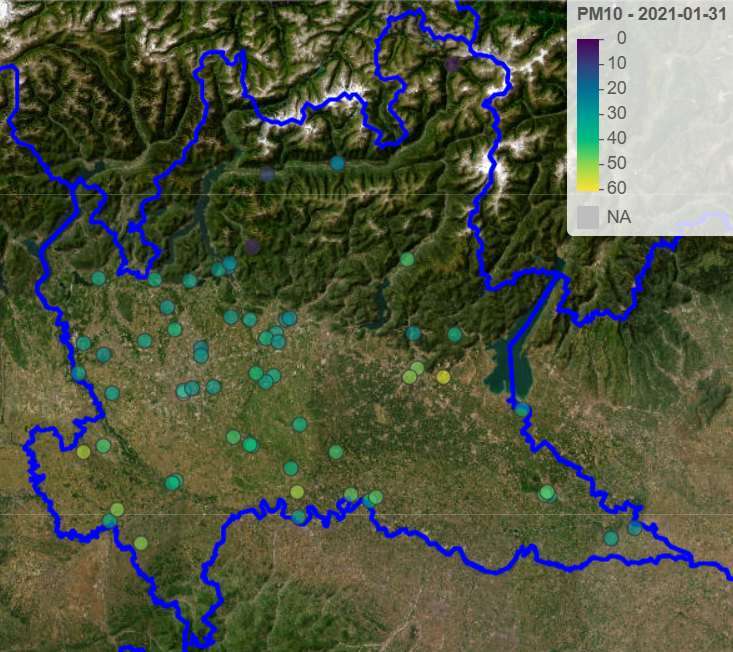}
\caption{January 31th, 2021. PM10}
\end{subfigure}
\hspace{2mm} 
\begin{subfigure}{0.49\textwidth}
\centering
\includegraphics[width=\textwidth]{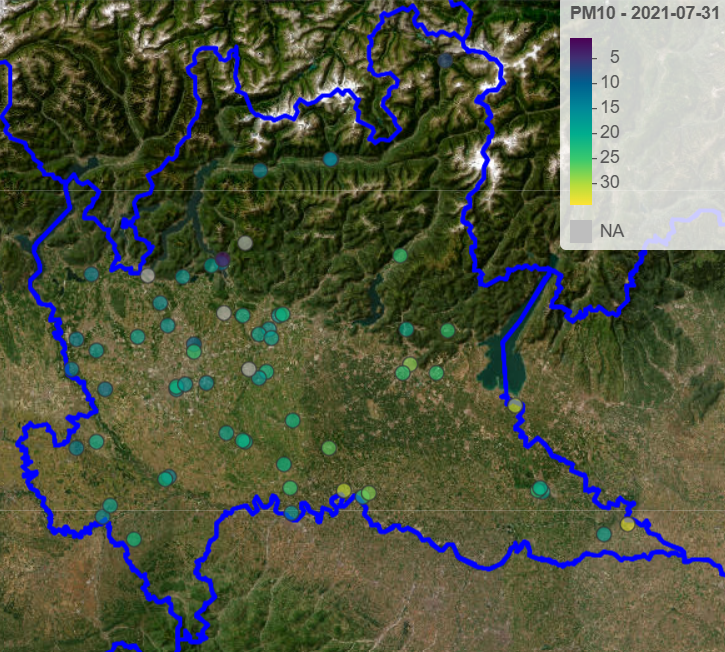}
\caption{July 31th, 2021. PM10}	
\end{subfigure}
\caption{Location of the 64 selected stations retrieving PM10 in Lombardy region.}
\label{fig:map}
\end{figure}

\begin{figure}[htbp!]
\centering
\includegraphics[angle=90,height=0.95\textheight]{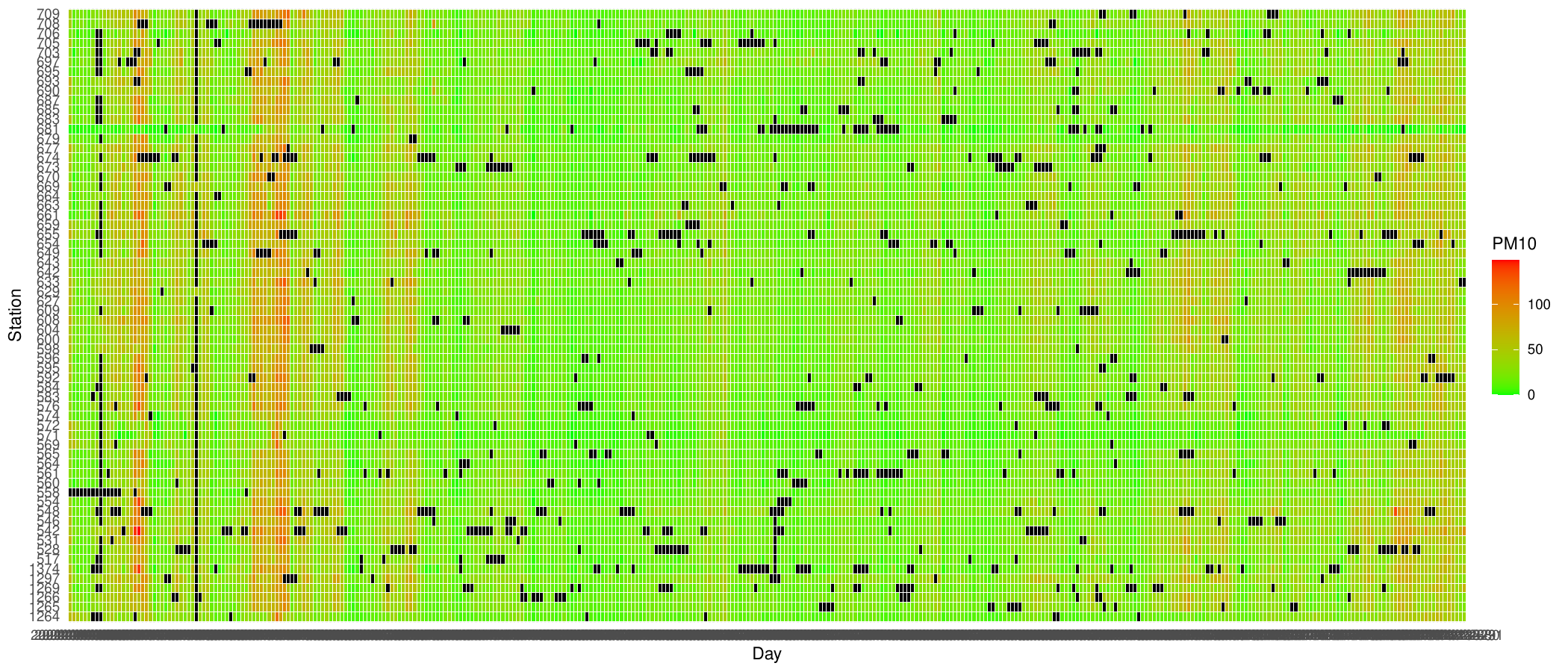}
\caption{PM$_{10}$ heatmap. Red indicates high values, green indicates low values, and missing data are shown in black.}
\label{fig:heatmap}
\end{figure}

From the full dataset, we extract observations corresponding to particulate matter with aerodynamic diameter smaller than 10 microns (PM$_{10}$) recorded by ground monitoring stations in the Lombardy region over the period January--December 2021.
We then identify the set of monitoring stations in Lombardy region collecting PM$_{10}$ series, discarding those with entirely missing measurements. 
Let $S$ denote the resulting set of stations ($|S|=64$), as reported in Figure \ref{fig:map}. 
For each $i \in S$, let $M_{i,t}$ denote the PM$_{10}$ concentration recorded at station $i$ on day $t$ (where $t$ indexes daily observations from 1 January to 31 December 2021). We stack the data into a station--day matrix:
\[
M
=
\big(M_{i,t}\big)_{i \in S,\, t=1,\ldots,T}
\in \mathbb{R}^{|S|\times T},
\qquad T=365,
\]
obtained by pivoting the long format panel into a wide matrix with stations as rows and days as columns. 
The resulting matrix contains observed concentration values and missing entries where stations did not report a measurement for a given day.
To visualize the missingness structure, we highlight missing values in black in the heatmap in Figure \ref{fig:heatmap}. 
The percentage of missing entries in this matrix is about 4\%. 

\begin{figure}[htbp!]
\centering
\includegraphics[width=0.6\textwidth]{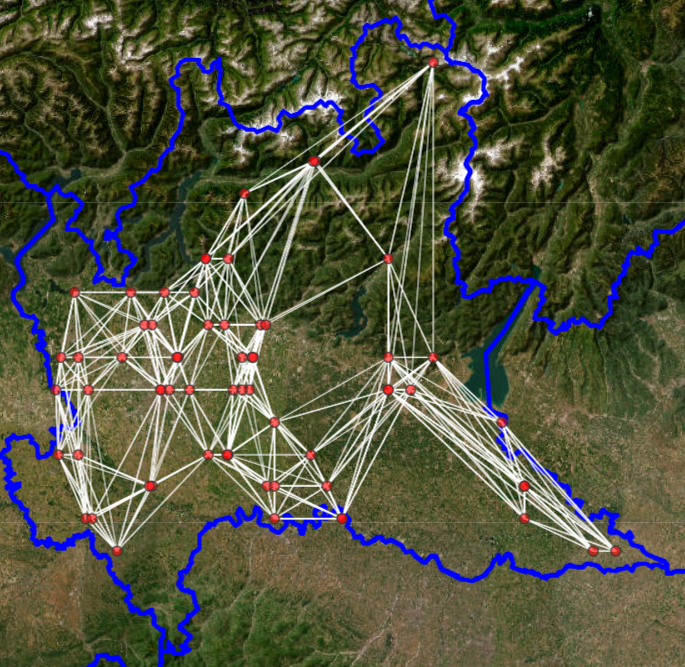}
\caption{K-nearest neighbours, using k = 10.}\label{fig:knn}	
\end{figure}

\subsection{Empirical strategy}

In this application, several empirical choices are required for the construction of the spatial filters and the implementation of the proposed method.

First, the tolerance level for the eigenvalues (the $\delta$ in Equation \ref{eq:retained}) is set to $\delta = 10^{-6}$ in order to discard numerically negligible components. 
Second, the number of spatial eigenvectors included in the matrix $A$ is determined by requiring that they jointly explain at least $90\%$ ($\tau = 0.90$) of the total spatial autocorrelation captured by the full set of eigenvectors $A_{\text{full}}$. 
This criterion ensures a parsimonious representation of spatial dependence while retaining the most relevant information.
The spatial weights matrix is constructed using a $k$-nearest neighbors structure with $k = 10$, as depicted in Figure \ref{fig:knn}. 
Alternative specifications, such as distance-based thresholds or spatial bandwidth criteria, could also be considered and may be explored in future work.

To assess the presence of spatial dependence in the original data, we compute Moran’s $I$ statistic. 
The results confirm significant spatial autocorrelation. 
For instance, on January 31, 2021, Moran’s $I$ equals $0.361$ (p-value $2.736 \times 10^{-16}$), while on July 31, 2021, it increases to $0.375$ (p-value $2.2 \times 10^{-16}$). 
These findings provide strong empirical motivation for incorporating spatial structure into matrix completion.

\begin{figure}[htbp!]
\begin{subfigure}{0.325\textwidth}
\centering
\includegraphics[width=\textwidth]{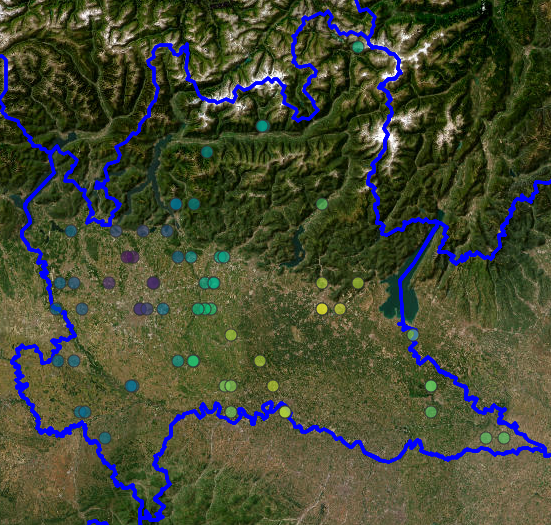}
\caption{First eigenvector}	
\end{subfigure}
\begin{subfigure}{0.318\textwidth}
\centering
\includegraphics[width=\textwidth]{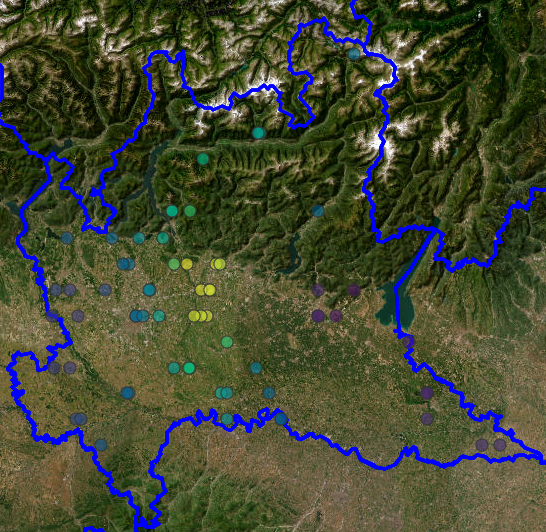}
\caption{Second eigenvector}	
\end{subfigure}
\begin{subfigure}{0.3255\textwidth}
\centering
\includegraphics[width=\textwidth]{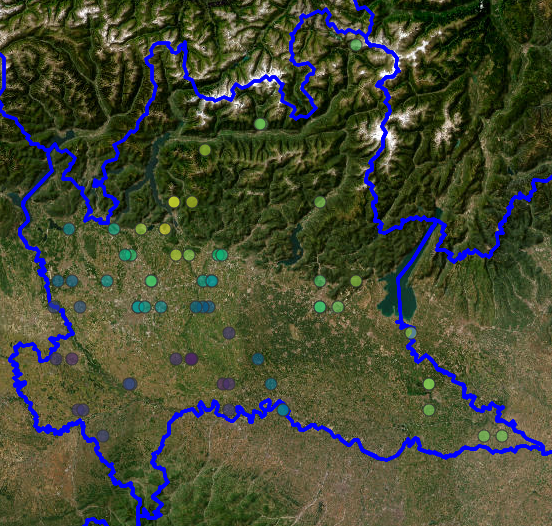}
\caption{Third eigenvector}	
\end{subfigure}

\begin{subfigure}{0.32\textwidth}
\centering
\includegraphics[width=\textwidth]{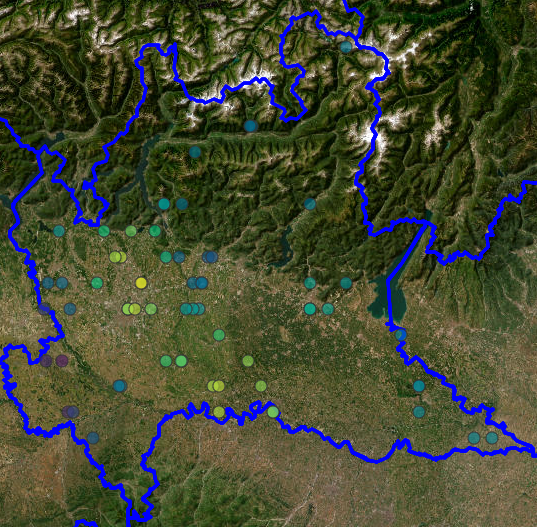}
\caption{Fourth eigenvector}	
\end{subfigure}
\begin{subfigure}{0.3325\textwidth}
\centering
\includegraphics[width=\textwidth]{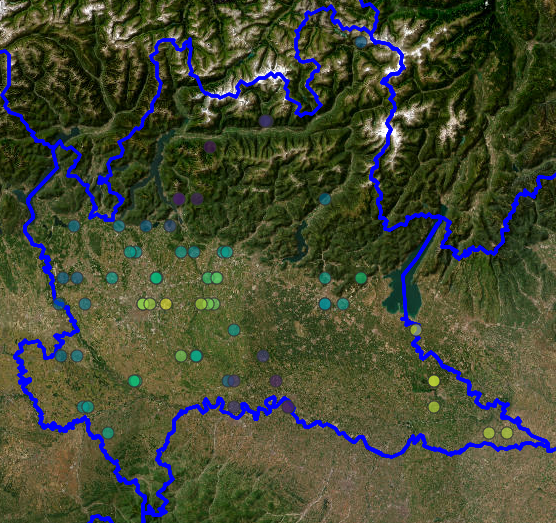}
\caption{Fifth eigenvector}	
\end{subfigure}
\begin{subfigure}{0.3265\textwidth}
\centering
\includegraphics[width=\textwidth]{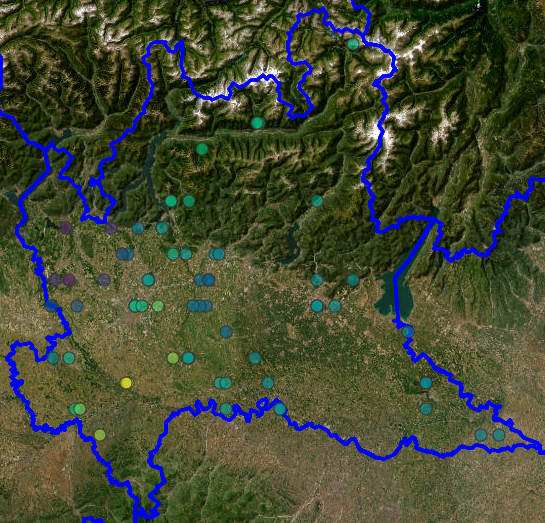}
\caption{Sixth eigenvector}	
\end{subfigure}

\begin{subfigure}{0.32\textwidth}
\centering
\includegraphics[width=\textwidth]{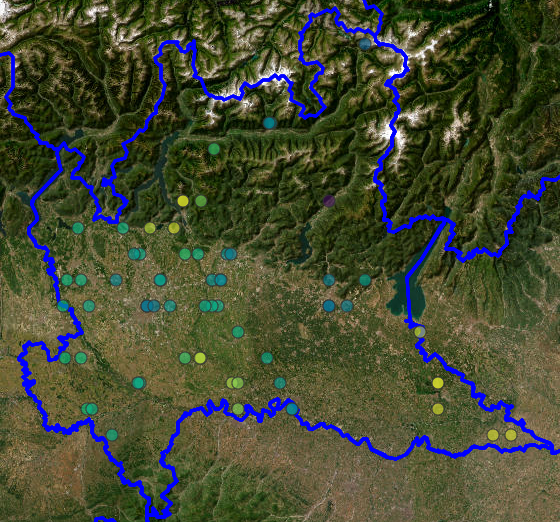}
\caption{Seventh eigenvector}	
\end{subfigure}
\caption{The seven chosen eigenvectors, ordered by their Moran's I index, computed using the W matrix based on the 10 nearest neighbours.}
\label{fig:eigen}
\end{figure}

Using the spatial weight matrix \(W\) defined by the \(k\)-nearest neighbors criterion with \(k = 10\), and applying the selection procedure described in Section \ref{sec:methods}, a total of seven eigenvectors were selected to be used as spatial filters. 
These eigenvectors are associated with the following values of Moran’s \(I\): $[0.925,\; 0.845,\; 0.785,\; 0.669,\; 0.571,\; 0.519,\; 0.374]$.
Figure \ref{fig:eigen} gives a representation of the seven chosen eigenvectors. 
The first eigenvector captures a broad regional gradient, while the subsequent eigenvectors describe increasingly localized clusters of stations exhibiting similar spatial behaviour. 
Together, they provide a multiscale representation of spatial dependence.
A key advantage of the ESF approach is that spatial autocorrelation is summarized through a small number of orthogonal components rather than a separate parameter for each station. 
In the Lombardy application, only seven eigenvectors explain approximately 90\% of the positive spatial autocorrelation signal, replacing 64 unrestricted unit effects.

Beyond improving imputation accuracy, these eigenvectors offer an interpretable description of the spatial organization of PM$_{10}$ concentrations, highlighting the main geographical structures driving air pollution variability. 
Estimated coefficients associated with the selected eigenvectors are $\hat{\alpha}=[11.87,\,-6.84,\,-11.41,\,
8.58,\,9.74,\,1.75,\,-1.90]$. 
Their magnitudes suggest that the first, third, fourth, and fifth eigenvectors provide the largest contribution to the spatial component of the model, implying that both broad regional gradients and intermediate-scale spatial clusters play an important role in explaining PM$_{10}$
variability across monitoring stations and, consequently, contribute to the reconstruction of the original matrix.

Computation times were measured on the same system used in the simulation study, equipped with an 11th Gen Intel Core i5-1135G7 processor (2.40 GHz) and 16 GB of RAM. Estimation of the ESFNNMC model on the Lombardy PM10 dataset required approximately 8.016 seconds, confirming the practical feasibility of the proposed approach on real-world environmental data.

It should be noted that the dataset considered in the empirical application is somewhat larger than those examined in the simulation study. However, from a practical perspective, the method could be implemented separately for each month, or alternatively on data aggregated at the monthly level, provided that this degree of temporal aggregation is adequate for the objectives of the analysis.

\subsection{Findings}

\begin{figure}[htbp!]
\centering
\includegraphics[width=1\textwidth]{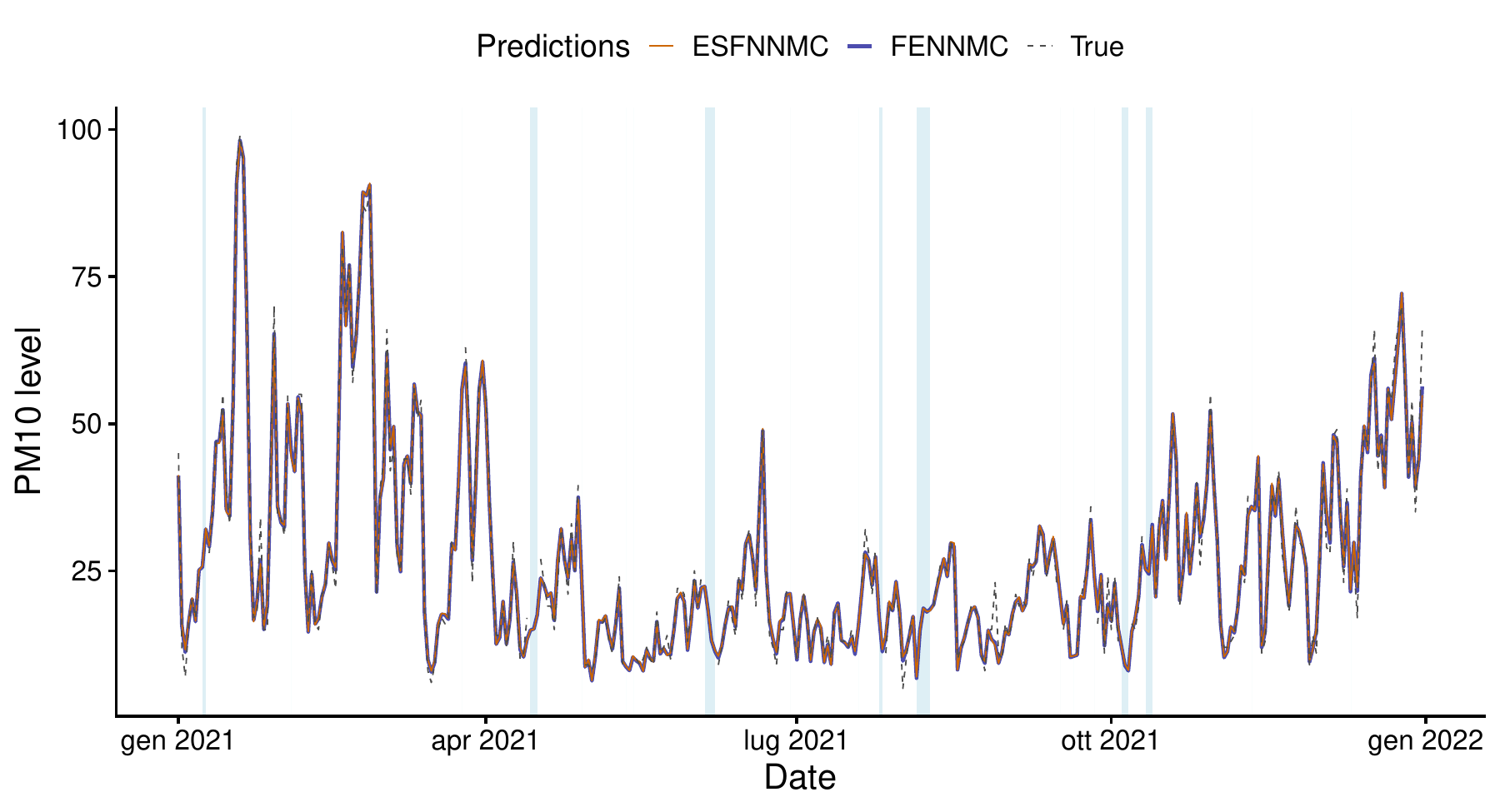}
\caption{Observed and reconstructed PM$_{10}$ values for the Dalmine Via Verdi station (province of Bergamo, Italy; 45.6497$^\circ$ N, 9.601223$^\circ$ E; elevation: 207\,m). Daily data for the year 2021.
ESFNNMC (in dark orange) accurately reproduces both seasonal dynamics and short-term peaks.}\label{fig:res}	
\end{figure}

Figure~\ref{fig:res} compares the observed PM$_{10}$ concentrations with the predictions obtained from ESFNNMC and FENNMC for the monitoring station located in Dalmine (Via Verdi) during 2021\footnote{In this application, we decided not to consider the version of the two methods that does not account for time fixed effects, as simulation results indicate that their performance is generally inferior.}. 
Both methods successfully reproduce the main temporal dynamics of the true series (dashed black line), including the pronounced winter peaks and the lower concentration levels observed during the summer months.

The predictions generated by the two approaches are generally very similar, indicating that both methods identify a comparable latent spatio-temporal structure. 
However, ESFNNMC tends to produce a slightly smoother trajectory, particularly during periods characterized by abrupt fluctuations. 
This behavior is consistent with the inclusion of eigenvector spatial filters, which borrow information from neighboring stations through the spatial dependence structure and therefore reduce the influence of local idiosyncratic variations.

The largest discrepancies between the models emerge around some of the highest PM$_{10}$ episodes observed during the winter season, where both methods slightly underestimate the magnitude of extreme peaks. 
However, the estimates obtained using the two methods are generally very similar, with a mean absolute difference of 0.220 and a maximum absolute difference of 1.497.
This result is expected, as low-rank matrix completion methods are primarily designed to recover the dominant latent structure rather than highly localized events. 
Nevertheless, the overall agreement between observed and predicted values remains satisfactory throughout the year.

\subsection{Validation strategy}

To provide an out-of-sample validation on real PM$_{10}$ data, we artificially masked a random subset of the originally observed entries. 
For each masking level, equal to 5\%, 10\%, and 20\%, we repeated the procedure over $B=200$ replications. 
In each replication, the models were estimated using the remaining observed cells, while predictive accuracy was evaluated only on the artificially obscured entries, for which the true values are known. 
This strategy preserves the original missingness pattern of the data while enabling a direct comparison between FENNMC and ESFNNMC with respect to out-of-sample reconstruction accuracy.

\begin{table}[ht]
\centering
\caption{Out-of-sample validation results based on artificially masking different levels of observed entries. 
Reported values correspond to the first quartile (Q1), median, and third quartile (Q3) of the MAPE on the validaton set over $B=200$ replications.}
\label{tab:validation_realdata}
\begin{tabular}{llccc}
\toprule
Masking & Method & Q1 & Median & Q3 \\
\midrule
\multirow{2}{*}{5\%}
& ESFNNMC & 16.26 & 16.96 & 17.93 \\
& FENNMC    & 15.61 & 16.36 & 17.19 \\
\midrule
\multirow{2}{*}{10\%}
& ESFNNMC & 16.71 & 17.31 & 17.91 \\
& FENNMC    & 16.12 & 16.65 & 17.28 \\
\midrule
\multirow{2}{*}{20\%}
& ESFNNMC & 17.47 & 17.85 & 18.36 \\
& FENNMC    & 16.91 & 17.35 & 17.74 \\
\bottomrule
\end{tabular}
\end{table}

The out-of-sample validation results reported in Table \ref{tab:validation_realdata} indicate that both methods achieve very similar levels of predictive accuracy. 
Although FENNMC attains slightly lower MAPE values across all masking levels, the differences remain modest in practical terms. 
This finding suggests that, for the Lombardy PM$_{10}$ dataset, the dominant variation is captured by temporal and low-rank components, leaving limited room for additional gains from spatial filtering. 
To further investigate this aspect, we computed the relative contribution of each model component based on its variance across elements. 
For ESFNNMC, temporal effects account for approximately 82.9\% of the reconstructed variability, compared with 14.7\% for the low-rank component and 2.5\% for the spatial component. 
Similarly, in FENNMC, temporal effects account for 80.5\% of the reconstructed variability, while the low-rank and unit-specific components contribute 11.7\% and 7.8\%, respectively. 
These results indicate that temporal heterogeneity represents the dominant source of variation in the dataset and help explain why the additional spatial information introduced through eigenvector spatial filters does not translate into substantial gains in predictive performance, although neither does it lead to a marked deterioration in accuracy. 
They also suggest that the spatial restriction imposed by ESFNNMC alters the decomposition of heterogeneity across model components: part of the variability captured by unrestricted unit effects in FENNMC appears to be absorbed by the low-rank component in ESFNNMC, leading to a more compact representation of unit-specific heterogeneity.

It should be emphasized that the objective of ESFNNMC is not solely predictive accuracy. By replacing unrestricted unit effects with a small set of spatial filters, the method provides a parsimonious and interpretable representation of spatial autocorrelation. In the Lombardy application, seven eigenvectors summarize the dominant spatial dependence structure, replacing 64 station-specific effects while maintaining a level of predictive accuracy comparable to that of FENNMC. Moreover, the estimated spatial component provides a direct characterization of the principal geographical patterns underlying PM$_{10}$ concentrations, an interpretation that is not available from unrestricted unit fixed effects.

\section{Conclusions} \label{sec:conclusion}

This paper introduces ESFNNMC, a spatial extension of nuclear norm matrix completion that incorporates eigenvector spatial filters to account for spatial dependence in partially observed spatio-temporal matrices. 
By replacing unit fixed effects with a parsimonious set of spatial eigenvectors, the proposed approach combines low-rank matrix completion and spatial statistical modeling while preserving computational efficiency through a block-coordinate descent estimation algorithm.

The simulation results show that ESFNNMC generally improves imputation accuracy compared with standard fixed-effects matrix completion, particularly in settings with moderate or strong spatial autocorrelation. 
The gains are especially evident when the number of spatial units is large relative to the temporal dimension. 
At the same time, the method achieves these improvements without a substantial increase in computational cost with respect to considered benchmarks.

Unlike unrestricted unit fixed effects, the spatial component estimated through Moran eigenvectors is directly interpretable, allowing the analyst to identify the dominant geographical structures contributing to the reconstruction. 
Thus, ESFNNMC should be viewed not only as an imputation method but also as a tool for characterizing spatial dependence in incomplete spatio-temporal datasets.

The empirical application to PM$_{10}$ concentrations measured at monitoring stations in Lombardy highlights the practical value of explicitly incorporating spatial information. 
Beyond reconstructing missing observations, the selected spatial filters provide an interpretable characterization of the underlying spatial organization of air pollution. 
In particular, the leading Moran eigenvectors capture broad geographical gradients, while subsequent eigenvectors describe increasingly localized spatial structures, thereby offering a multiscale representation of the spatial dependence governing PM$_{10}$ concentrations across monitoring stations.

Several directions for future research deserve attention. 
First, the current specification relies on time fixed effects; a natural extension would replace them with an explicit temporal dependence structure, such as autoregressive or state-space dynamics \citep{rodeschini2026multivariate}. 
Second, the present study assumes a Missing At Random (MAR) mechanism, as commonly done in matrix completion settings. 
However, environmental and emissions data may exhibit Missing Not at Random (MNAR) mechanisms, whereby the probability of missingness depends on the unobserved values themselves. 
Recent extensions of low-rank matrix completion explicitly address MNAR processes through propensity weighting, joint modeling of the outcome and missingness mechanisms, or causal approaches \citep{Sportisse2018Imputation,ma2019missing,Choi2023Matrix}. 
Extending ESFNNMC to accommodate such missingness mechanisms therefore represents an important direction for future research.
Third, alternative definitions of the spatial weights matrix, including distance-based and adaptive neighborhood structures, may further improve performance. 
The implementation of time-varying weight matrices could also be adopted, since spatial interaction patterns can evolve over time. 
Future developments may additionally allow the inclusion of explanatory features within the matrix completion framework, following, for example, \cite{mao2019matrix}.
From a theoretical perspective, establishing formal results on identifiability, consistency, finite-sample approximation error, and statistical recovery guarantees for the proposed estimator represents another important direction for future research.
Finally, the role of negatively autocorrelated eigenvectors remains an open topic for investigation.

Although the present work focuses on missing-data reconstruction, the proposed framework can also be naturally extended to counterfactual analysis and policy evaluation. 
By masking outcomes associated with treated units during the post-intervention period, ESFNNMC can be used to estimate the untreated potential outcomes and therefore quantify the effects of environmental, transportation, or emissions-related policies while explicitly accounting for spatial dependence.

Overall, ESFNNMC provides a flexible and computationally efficient framework for reconstructing incomplete spatio-temporal datasets and represents a bridge between matrix completion methods, spatial statistics, and counterfactual policy analysis.

\section*{Disclosure Statement}

The author reports that there are no competing interests to declare.

\section*{Funding}
This work was supported by the project "Study of mobile phone siGNals for the evalUation of the interconnections between Mobility and the environment in Lombardia (SIGNUM)" CUP: F53D23010910001- PRIN 2022 PNRR M4C2 - financed by the European Union - Next Generation EU (DD MUR n. 1409 del 14/09/2022).

\section*{Data and Code Availability Statement}
The data and the code implementing the proposed method, reproducing the simulation study and empirical application presented in this paper, are freely available at the following GITHUB repository: \url{https://github.com/rodolfometulini/ESFNNMC}.

\section*{Declaration of Generative AI Use}

Generative artificial intelligence tools, including ChatGPT (OpenAI), were
used during the preparation of the manuscript solely to improve the clarity,
readability, and English language of the text. The author critically reviewed and edited all AI-assisted content and takes full responsibility for the final manuscript. 
Generative AI was not used for the implementation of the study, statistical analysis, or interpretation of the results.

\section*{Acknowledgments}
The author would like to express his gratitude to Roberto Patuelli, Francesco Biancalani, and the attendees of the CFE-CMStatistics 2025 conference and METMA XII workshop for useful feedback.

\bibliographystyle{plainnat}
\bibliography{refs}

\clearpage
\appendix
\section{Supplementary Simulation Results}\label{app:simulation_tables}
\renewcommand{\thetable}{A\arabic{table}}
\setcounter{table}{0}

This appendix reports supplementary results from the simulation study that complement the comparison between ESFNNMC and FENNMC based on imputation accuracy presented in the main text. 
In particular, we summarize the values of the optimal regularization parameter $\lambda$ selected by cross-validation (Tables \ref{tab:lambda_rank5_10x10}, \ref{tab:lambda_rank5_10x50}, \ref{tab:lambda_rank10_50x10}, and \ref{tab:lambda_rank10_30x30}) and the number of retained Moran eigenvectors used as spatial filters in ESFNNMC (Tables \ref{tab:filters_rank5_10x50}, \ref{tab:filters_rank10_50x10}, and \ref{tab:filters_rank10_30x30}) for each simulation setting.

The optimal value of $\lambda$ provides information on the amount of nuclear-norm regularization required by the competing methods. 
Across all scenarios, the selected values generally increase with the percentage of missing observations, indicating that stronger shrinkage is needed as the amount of available information decreases. 
Moreover, ESFNNMC typically selects values of $\lambda$ that are comparable to, or slightly smaller than, those selected by FENNMC. 
This suggests that explicitly accounting for spatially structured heterogeneity through Moran eigenvectors reduces the amount of regularization required to recover the latent low-rank component.

The appendix also reports the number of spatial filters retained by the proposed selection procedure. 
These results show that the number of selected eigenvectors remains remarkably stable across different levels of missingness and simulation settings. 
Even for the largest matrices, only a relatively small subset of Moran eigenvectors is retained, demonstrating that ESFNNMC provides a parsimonious representation of spatial heterogeneity. 
In all configurations, the selected filters correspond to only a small fraction of the total number of spatial units, while preserving the predictive performance as reported in the main text. 

These findings support the effectiveness of the proposed filters' selection strategy in achieving a favorable trade-off between model complexity and reconstruction accuracy.

\begin{table}[!htbp]
\centering
\caption{Median selected $\lambda$ across methods for the $10\times 10$ matrix with rank $5$ and three regimes.}
\label{tab:lambda_rank5_10x10}
\begin{adjustbox}{max totalheight=0.97\textheight,max width=\textwidth}
\begin{tabular}{lcccc}
\toprule
Missing (\%) & FENNMC (FE) & ESFNNMC (FE) & FENNMC (No FE) & ESFNNMC (No FE) \\
\midrule
\multicolumn{5}{c}{\textbf{Panel A: $\rho=0$, $\phi=0$}} \\
\midrule
2  & 0.2640 & 0.2310 & 0.4444 & 0.13444 \\
4  & 0.3614 & 0.2525 & 0.6188 & 0.13614 \\
6  & 0.3578 & 0.2607 & 0.5444 & 0.10288 \\
8  & 0.4392 & 0.3258 & 0.6260 & 0.10676 \\
10 & 0.4538 & 0.3455 & 0.6729 & 0.16331 \\
15 & 0.5772 & 0.4248 & 0.7092 & 0.23998 \\
20 & 0.8956 & 0.5861 & 1.0030 & 0.19250 \\
25 & 0.8406 & 0.9178 & 1.2353 & 0.24832 \\
\midrule
\multicolumn{5}{c}{\textbf{Panel B: $\rho=0$, $\phi=0.4$}} \\
\midrule
2  & 0.2517 & 0.2525 & 0.3579 & 0.13127 \\
4  & 0.2634 & 0.2996 & 0.4245 & 0.08495 \\
6  & 0.3073 & 0.2845 & 0.5124 & 0.13875 \\
8  & 0.3668 & 0.3472 & 0.5328 & 0.13809 \\
10 & 0.4060 & 0.4063 & 0.4500 & 0.16379 \\
15 & 0.6518 & 0.5827 & 0.6414 & 0.21018 \\
20 & 0.8045 & 0.6719 & 0.9921 & 0.21649 \\
25 & 0.9405 & 0.8907 & 1.0523 & 0.20887 \\
\midrule
\multicolumn{5}{c}{\textbf{Panel C: $\rho=0$, $\phi=0.8$}} \\
\midrule
2  & 0.2530 & 0.2801 & 0.2386 & 0.08101 \\
4  & 0.2839 & 0.3238 & 0.2396 & 0.11085 \\
6  & 0.3614 & 0.3159 & 0.2957 & 0.09476 \\
8  & 0.3170 & 0.3774 & 0.3321 & 0.07251 \\
10 & 0.4214 & 0.4538 & 0.4132 & 0.10519 \\
15 & 0.6426 & 0.5308 & 0.4178 & 0.09605 \\
20 & 0.5564 & 0.6694 & 0.7877 & 0.12698 \\
25 & 0.9305 & 1.0589 & 0.7873 & 0.12738 \\
\midrule
\multicolumn{5}{c}{\textbf{Panel D: $\rho=0.4$, $\phi=0$}} \\
\midrule
2  & 0.2719 & 0.2350 & 0.4425 & 0.07698 \\
4  & 0.3808 & 0.3184 & 0.4660 & 0.17968 \\
6  & 0.3579 & 0.2612 & 0.4356 & 0.11659 \\
8  & 0.4009 & 0.3199 & 0.5268 & 0.13523 \\
10 & 0.3663 & 0.4235 & 0.5098 & 0.11089 \\
15 & 0.4889 & 0.4320 & 0.6589 & 0.22152 \\
20 & 0.8205 & 0.7027 & 0.9434 & 0.17507 \\
25 & 0.8142 & 0.9186 & 1.1711 & 0.27150 \\
\midrule
\multicolumn{5}{c}{\textbf{Panel E: $\rho=0.4$, $\phi=0.4$}} \\
\midrule
2  & 0.2782 & 0.2751 & 0.2980 & 0.11747 \\
4  & 0.2643 & 0.3486 & 0.3798 & 0.10079 \\
6  & 0.3445 & 0.3473 & 0.3871 & 0.09341 \\
8  & 0.3701 & 0.3066 & 0.4433 & 0.13177 \\
10 & 0.3463 & 0.3990 & 0.4195 & 0.14750 \\
15 & 0.6014 & 0.6082 & 0.6631 & 0.16921 \\
20 & 0.8359 & 0.7313 & 0.8677 & 0.25254 \\
25 & 0.9559 & 1.0086 & 0.9461 & 0.15085 \\
\midrule
\multicolumn{5}{c}{\textbf{Panel F: $\rho=0.4$, $\phi=0.8$}} \\
\midrule
2  & 0.2041 & 0.2609 & 0.1642 & 0.07571 \\
4  & 0.2468 & 0.3718 & 0.1896 & 0.08514 \\
6  & 0.3763 & 0.3506 & 0.2822 & 0.08619 \\
8  & 0.3287 & 0.4315 & 0.2672 & 0.07234 \\
10 & 0.3923 & 0.5134 & 0.2992 & 0.08795 \\
15 & 0.5516 & 0.6448 & 0.4030 & 0.11794 \\
20 & 0.6764 & 0.6912 & 0.6661 & 0.18922 \\
25 & 1.0321 & 1.2428 & 0.7666 & 0.09814 \\
\midrule
\multicolumn{5}{c}{\textbf{Panel G: $\rho=0.8$, $\phi=0$}} \\
\midrule
2  & 0.3174 & 0.2450 & 0.2119 & 0.09071 \\
4  & 0.2816 & 0.2909 & 0.2829 & 0.17254 \\
6  & 0.3590 & 0.3131 & 0.2502 & 0.07331 \\
8  & 0.3078 & 0.3267 & 0.3199 & 0.08734 \\
10 & 0.4011 & 0.3715 & 0.2971 & 0.12432 \\
15 & 0.5603 & 0.4743 & 0.4604 & 0.14479 \\
20 & 0.8622 & 0.7053 & 0.5484 & 0.19597 \\
25 & 1.0967 & 0.8865 & 0.8368 & 0.24865 \\
\midrule
\multicolumn{5}{c}{\textbf{Panel H: $\rho=0.8$, $\phi=0.4$}} \\
\midrule
2  & 0.2321 & 0.2764 & 0.1728 & 0.09094 \\
4  & 0.2939 & 0.3441 & 0.2146 & 0.12647 \\
6  & 0.3050 & 0.3419 & 0.1801 & 0.09651 \\
8  & 0.3343 & 0.3670 & 0.2394 & 0.11374 \\
10 & 0.4008 & 0.3692 & 0.1826 & 0.11506 \\
15 & 0.6132 & 0.6153 & 0.2658 & 0.14634 \\
20 & 0.7727 & 0.7211 & 0.6333 & 0.25142 \\
25 & 1.1684 & 1.0546 & 0.7571 & 0.17869 \\
\midrule
\multicolumn{5}{c}{\textbf{Panel I: $\rho=0.8$, $\phi=0.8$}} \\
\midrule
2  & 0.1538 & 0.2764 & 0.1728 & 0.09094 \\
4  & 0.2356 & 0.3441 & 0.2146 & 0.12647 \\
6  & 0.3157 & 0.3419 & 0.1801 & 0.09651 \\
8  & 0.3080 & 0.3670 & 0.2394 & 0.11374 \\
10 & 0.3692 & 0.3692 & 0.1826 & 0.11506 \\
15 & 0.6132 & 0.6153 & 0.2658 & 0.14634 \\
20 & 0.7727 & 0.7211 & 0.6333 & 0.25142 \\
25 & 1.1684 & 1.0546 & 0.7571 & 0.17869 \\
\bottomrule
\end{tabular}
\end{adjustbox}
\end{table}

\begin{table}[!htbp]
\centering
\caption{Median selected $\lambda$ (10$\times$50 matrix, rank 5, three regimes).}
\label{tab:lambda_rank5_10x50}
\begin{tabular}{lcccc}
\toprule
 & FENNMC (FE) & ESFNNMC (FE) & FENNMC (No FE) & ESFNNMC (No FE) \\
\midrule
All panels & 0 & 0 & 0 & 0 \\
\bottomrule
\end{tabular}
\end{table}

\begin{table}[!htbp]
\centering
\caption{Median number of spatial filters (10$\times$50 matrix, rank 5, three regimes).}
\label{tab:filters_rank5_10x50}
\begin{tabular}{lc}
\toprule
Missing (\%) & Filters \\
\midrule
2--25 & 2 \\
\bottomrule
\end{tabular}
\end{table}

\begin{table}[!htbp]
\centering
\caption{Median optimal $\lambda$ across methods (50$\times$10 matrix, rank 10, three regimes).}
\label{tab:lambda_rank10_50x10}
\begin{adjustbox}{max totalheight=0.97\textheight,max width=\textwidth}
\begin{tabular}{lcccc}
\toprule
 & FENNMC (FE) & ESFNNMC (FE) & FENNMC (No FE) & ESFNNMC (No FE) \\
\midrule

\multicolumn{5}{c}{\textbf{Panel A: $\rho=0$, $\phi=0$}} \\
\midrule
2  & 0 & 0 & 0 & 0 \\
4  & 0 & 0 & 0 & 0 \\
6  & 0 & 0 & 0 & 0 \\
8  & 0 & 0 & 0 & 0 \\
10 & 0 & 0.004642 & 0.005068 & 0 \\
15 & 0.1447 & 0.063932 & 0.176328 & 0 \\
20 & 0.1645 & 0.111785 & 0.244733 & 0.01578 \\
25 & 0.2312 & 0.222384 & 0.328464 & 0.02132 \\

\midrule
\multicolumn{5}{c}{\textbf{Panel B: $\rho=0$, $\phi=0.4$}} \\
\midrule
2  & 0 & 0 & 0 & 0 \\
4  & 0 & 0 & 0 & 0 \\
6  & 0 & 0 & 0 & 0 \\
8  & 0 & 0 & 0 & 0 \\
10 & 0 & 0.004479 & 0.004372 & 0 \\
15 & 0.1135 & 0.084538 & 0.142193 & 0 \\
20 & 0.1444 & 0.133148 & 0.200410 & 0 \\
25 & 0.1940 & 0.218743 & 0.239603 & 0 \\

\midrule
\multicolumn{5}{c}{\textbf{Panel C: $\rho=0$, $\phi=0.8$}} \\
\midrule
2  & 0 & 0 & 0 & 0 \\
4  & 0 & 0 & 0 & 0 \\
6  & 0 & 0 & 0 & 0 \\
8  & 0 & 0 & 0 & 0 \\
10 & 0 & 0 & 0 & 0 \\
15 & 0.01059 & 0.006203 & 0.08321 & 0 \\
20 & 0.11497 & 0.014034 & 0.11255 & 0 \\
25 & 0.17960 & 0.117704 & 0.18605 & 0 \\

\midrule
\multicolumn{5}{c}{\textbf{Panel D: $\rho=0.4$, $\phi=0$}} \\
\midrule
2  & 0 & 0 & 0 & 0 \\
4  & 0 & 0 & 0 & 0 \\
6  & 0 & 0 & 0 & 0 \\
8  & 0 & 0 & 0 & 0 \\
10 & 0 & 0.004906 & 0.004619 & 0 \\
15 & 0.1455 & 0.052494 & 0.166276 & 0 \\
20 & 0.1651 & 0.118948 & 0.206777 & 0 \\
25 & 0.2544 & 0.224797 & 0.320612 & 0 \\

\midrule
\multicolumn{5}{c}{\textbf{Panel E: $\rho=0.4$, $\phi=0.4$}} \\
\midrule
2  & 0 & 0 & 0 & 0 \\
4  & 0 & 0 & 0 & 0 \\
6  & 0 & 0 & 0 & 0 \\
8  & 0 & 0 & 0 & 0 \\
10 & 0 & 0.004674 & 0 & 0 \\
15 & 0.1016 & 0.102328 & 0.1162 & 0 \\
20 & 0.1453 & 0.140551 & 0.1926 & 0 \\
25 & 0.2005 & 0.230497 & 0.2381 & 0 \\

\midrule
\multicolumn{5}{c}{\textbf{Panel F: $\rho=0.4$, $\phi=0.8$}} \\
\midrule
2  & 0 & 0 & 0 & 0 \\
4  & 0 & 0 & 0 & 0 \\
6  & 0 & 0 & 0 & 0 \\
8  & 0 & 0 & 0 & 0 \\
10 & 0 & 0 & 0 & 0 \\
15 & 0.007622 & 0.006963 & 0.03352 & 0 \\
20 & 0.109440 & 0.062773 & 0.127870 & 0 \\
25 & 0.168075 & 0.117829 & 0.18464 & 0 \\

\midrule
\multicolumn{5}{c}{\textbf{Panel G: $\rho=0.8$, $\phi=0$}} \\
\midrule
2  & 0 & 0 & 0 & 0 \\
4  & 0 & 0 & 0 & 0 \\
6  & 0 & 0 & 0 & 0 \\
8  & 0 & 0.003817 & 0 & 0 \\
10 & 0 & 0.005145 & 0 & 0 \\
15 & 0.1426 & 0.083144 & 0.1049 & 0 \\
20 & 0.1608 & 0.120623 & 0.1478 & 0 \\
25 & 0.2693 & 0.225497 & 0.2184 & 0 \\

\midrule
\multicolumn{5}{c}{\textbf{Panel H: $\rho=0.8$, $\phi=0.4$}} \\
\midrule
2  & 0 & 0 & 0 & 0 \\
4  & 0 & 0 & 0 & 0 \\
6  & 0 & 0 & 0 & 0 \\
8  & 0 & 0 & 0 & 0 \\
10 & 0 & 0 & 0 & 0 \\
15 & 0.09218 & 0.119396 & 0.006353 & 0 \\
20 & 0.16944 & 0.160756 & 0.127870 & 0 \\
25 & 0.21119 & 0.248181 & 0.210778 & 0 \\

\midrule
\multicolumn{5}{c}{\textbf{Panel I: $\rho=0.8$, $\phi=0.8$}} \\
\midrule
2  & 0 & 0 & 0 & 0 \\
4  & 0 & 0 & 0 & 0 \\
6  & 0 & 0 & 0 & 0 \\
8  & 0 & 0 & 0 & 0 \\
10 & 0 & 0 & 0 & 0 \\
15 & 0.008341 & 0.00776 & 0 & 0 \\
20 & 0.100336 & 0.09471 & 0.04652 & 0 \\
25 & 0.179297 & 0.13039 & 0.14141 & 0 \\

\bottomrule
\end{tabular}
\end{adjustbox}
\end{table}

\begin{table}[!htbp]
\centering
\caption{Median number of spatial filters (50$\times$10 matrix, rank 10, three regimes).}
\label{tab:filters_rank10_50x10}
\begin{tabular}{lc}
\toprule
Missing (\%) & Filters \\
\midrule
2–25 & 13 \\
\bottomrule
\end{tabular}
\end{table}

\begin{table}[!htbp]
\centering
\caption{Median optimal $\lambda$ across methods (30$\times$30 matrix, rank 10, three regimes).}
\label{tab:lambda_rank10_30x30}
\begin{adjustbox}{max totalheight=0.97\textheight,max width=\textwidth}
\begin{tabular}{lcccc}
\toprule
 & FENNMC (FE) & ESFNNMC (FE) & FENNMC (No FE) & ESFNNMC (No FE) \\
\midrule

\multicolumn{5}{c}{\textbf{Panel A: $\rho=0$, $\phi=0$}} \\
\midrule
2  & 0 & 0 & 0 & 0 \\
4  & 0 & 0 & 0 & 0 \\
6  & 0 & 0 & 0 & 0 \\
8  & 0 & 0 & 0.003362 & 0 \\
10 & 0 & 0.002516 & 0.004139 & 0 \\
15 & 0.008626 & 0.004696 & 0.014874 & 0 \\
20 & 0.022016 & 0.015258 & 0.031152 & 0.01369 \\
25 & 0.033196 & 0.030737 & 0.050976 & 0.01810 \\

\midrule
\multicolumn{5}{c}{\textbf{Panel B: $\rho=0$, $\phi=0.4$}} \\
\midrule
2  & 0 & 0 & 0 & 0 \\
4  & 0 & 0 & 0 & 0 \\
6  & 0 & 0 & 0 & 0 \\
8  & 0 & 0 & 0.003232 & 0 \\
10 & 0 & 0 & 0.003486 & 0 \\
15 & 0.005158 & 0.007598 & 0.013222 & 0 \\
20 & 0.018746 & 0.017103 & 0.030588 & 0.01297 \\
25 & 0.037043 & 0.032368 & 0.046307 & 0.01922 \\

\midrule
\multicolumn{5}{c}{\textbf{Panel C: $\rho=0$, $\phi=0.8$}} \\
\midrule
2  & 0 & 0 & 0 & 0 \\
4  & 0 & 0 & 0 & 0 \\
6  & 0 & 0 & 0 & 0 \\
8  & 0 & 0 & 0 & 0 \\
10 & 0 & 0 & 0 & 0 \\
15 & 0.004236 & 0.006701 & 0.005447 & 0 \\
20 & 0.013317 & 0.018957 & 0.017428 & 0 \\
25 & 0.028619 & 0.037083 & 0.032371 & 0 \\

\midrule
\multicolumn{5}{c}{\textbf{Panel D: $\rho=0.4$, $\phi=0$}} \\
\midrule
2  & 0 & 0 & 0 & 0 \\
4  & 0 & 0 & 0 & 0 \\
6  & 0 & 0 & 0 & 0 \\
8  & 0 & 0 & 0 & 0 \\
10 & 0 & 0 & 0 & 0 \\
15 & 0.007027 & 0.004807 & 0.013568 & 0 \\
20 & 0.020916 & 0.016507 & 0.028250 & 0 \\
25 & 0.037852 & 0.030646 & 0.047474 & 0 \\

\midrule
\multicolumn{5}{c}{\textbf{Panel E: $\rho=0.4$, $\phi=0.4$}} \\
\midrule
2  & 0 & 0 & 0 & 0 \\
4  & 0 & 0 & 0 & 0 \\
6  & 0 & 0 & 0 & 0 \\
8  & 0 & 0 & 0 & 0 \\
10 & 0 & 0.003046 & 0 & 0 \\
15 & 0.005398 & 0.007731 & 0.004598 & 0 \\
20 & 0.017280 & 0.019284 & 0.015194 & 0 \\
25 & 0.037593 & 0.038637 & 0.044812 & 0 \\

\midrule
\multicolumn{5}{c}{\textbf{Panel F: $\rho=0.4$, $\phi=0.8$}} \\
\midrule
2  & 0 & 0 & 0 & 0 \\
4  & 0 & 0 & 0 & 0 \\
6  & 0 & 0 & 0 & 0 \\
8  & 0 & 0 & 0 & 0 \\
10 & 0 & 0 & 0 & 0 \\
15 & 0 & 0.006542 & 0 & 0 \\
20 & 0.01386 & 0.021497 & 0.015194 & 0 \\
25 & 0.02802 & 0.039286 & 0.028859 & 0 \\

\midrule
\multicolumn{5}{c}{\textbf{Panel G: $\rho=0.8$, $\phi=0$}} \\
\midrule
2  & 0 & 0 & 0 & 0 \\
4  & 0 & 0 & 0 & 0 \\
6  & 0 & 0 & 0 & 0 \\
8  & 0 & 0 & 0 & 0 \\
10 & 0 & 0.001353 & 0 & 0 \\
15 & 0.006487 & 0.005934 & 0.007444 & 0 \\
20 & 0.021899 & 0.016201 & 0.018697 & 0 \\
25 & 0.040237 & 0.030967 & 0.030563 & 0 \\

\midrule
\multicolumn{5}{c}{\textbf{Panel H: $\rho=0.8$, $\phi=0.4$}} \\
\midrule
2  & 0 & 0 & 0 & 0 \\
4  & 0 & 0 & 0 & 0 \\
6  & 0 & 0 & 0 & 0 \\
8  & 0 & 0 & 0 & 0 \\
10 & 0 & 0 & 0 & 0 \\
15 & 0.005720 & 0.007744 & 0.005215 & 0 \\
20 & 0.018447 & 0.019028 & 0.014653 & 0 \\
25 & 0.037739 & 0.041281 & 0.029277 & 0 \\

\midrule
\multicolumn{5}{c}{\textbf{Panel I: $\rho=0.8$, $\phi=0.8$}} \\
\midrule
2  & 0 & 0 & 0 & 0 \\
4  & 0 & 0 & 0 & 0 \\
6  & 0 & 0 & 0 & 0 \\
8  & 0 & 0 & 0 & 0 \\
10 & 0 & 0 & 0 & 0 \\
15 & 0 & 0.007252 & 0 & 0 \\
20 & 0.01310 & 0.019840 & 0.007222 & 0 \\
25 & 0.02562 & 0.039372 & 0.020297 & 0 \\

\bottomrule
\end{tabular}
\end{adjustbox}
\end{table}

\begin{table}[!htbp]
\centering
\caption{Median number of spatial filters (30$\times$30 matrix, rank 10, three regimes).}
\label{tab:filters_rank10_30x30}
\begin{tabular}{lc}
\toprule
Missing (\%) & Filters \\
\midrule
2–25 & 9 \\
\bottomrule
\end{tabular}
\end{table}

\end{document}